%% file: bcd.tex
\documentclass[useAMS,usenatbib]{mn2e}
\usepackage{graphicx, rotating}
\usepackage{aas_macros}  
\usepackage{graphicx,subfig}
\usepackage{amssymb}
\usepackage{rotating}
\usepackage{multirow}
\usepackage[breaklinks,colorlinks=true,citecolor=black,linkcolor=black,urlcolor=blue,bookmarks=true]{hyperref}
\usepackage{breakurl}  
 \usepackage{amsmath}

\newcommand{\kms}{km\,s$^{-1}$}
\newcommand{\reff}{R$_e$}
\newcommand{\hone}{H{\sc i}}

\newcommand{\halpha}{H$_\alpha$}

\newcommand{\sthree}{[S{\sc iii}]}
\newcommand{\ulyss}{\href{http://ulyss.univ-lyon1.fr}{\textcolor{black}{\emph{ULySS}}}}
\newcommand{\msol}{M$_\odot$}

\title[Stellar kinematics of 4 BCDs]{On the origin of bursts in blue
  compact dwarf galaxies: clues from kinematics and
  stellar populations} \author[Koleva et
al.]{M. Koleva$^{1}$\thanks{E-mail: mina.koleva@ugent.be (MK);
    sven.derijcke@ugent.be (SDR)}\thanks{Marie Curie/FWO Fellow},
  S. De Rijcke$^{1}$\footnotemark[1]\thanks{Thanks the Special
    Research Fund (BOF)},
W. W. Zeilinger$^{2}$,
R. Verbeke$^{1}$,
J. Schroyen$^{1}$\thanks{Thanks FWO}, 
\and L. Vermeylen$^{1}$\\
$^{1}$ Sterrenkundig Observatorium, Ghent University, Krijgslaan 281,
S9, 9000 Ghent, Belgium \\
$^{2}$ Institut f\"ur Astrophysik, Universit'\"at Wien
T\"urkenschanzstrasse 17, A-1180, Wien, Austria 
}
\begin{document}

\date{Accepted 201X XXXX XX. Received 201X XXXXX XX; in original form 201X XXXX XX}

\pagerange{\pageref{firstpage}--\pageref{lastpage}} \pubyear{201X}

\maketitle

\label{firstpage}

\begin{abstract}
  Blue compact dwarf galaxies (BCDs) form stars at, for their sizes,
  extraordinarily high rates. In this paper, we study what triggers
  this starburst and what is the fate of the galaxy once its gas fuel
  is exhausted. We select four BCDs with smooth outer regions,
  indicating them as possible progenitors of dwarf elliptical
  galaxies. We have obtained photometric and spectroscopic data with
  the FORS and ISAAC instruments on the VLT. We analyse their
  infra-red spectra using a full spectrum fitting technique which
  yields the kinematics of their stars and ionized gas together with
  their stellar population characteristics. We find that the
  \emph{stellar} velocity to velocity dispersion ratio
  ($(v/\sigma)_\star$) of our BCDs is of the order of 1.5, similar to
  that of dwarf elliptical galaxies. Thus, those objects do not
  require significant (if any) loss of angular momentum to fade into
  early type dwarfs. This finding is in discordance with previous
  studies, which however compared the stellar kinematics of dwarf
  elliptical galaxies with the gaseous kinematics of star forming
  dwarfs. The stellar velocity fields of our objects are very
  disturbed and the star-formation regions are often kinematically
  decoupled from the rest of the galaxy. These regions can be more or
  less metal rich with respect to the galactic body, and sometimes
  they are long lived.  These characteristics prevent us from
  pinpointing a unique trigger of the star formation, even within the
  same galaxy. Gas impacts, mergers, and in-spiraling gas clumps are
  all possible star-formation ignitors for our targets.
\end{abstract}


\section{Introduction}
One of the key questions occupying astrophysicists today is how
star formation proceeds in galaxies. What triggers star formation and
what ceases it?  The reason for this interest is twofold: firstly,
observations of stellar light carry most
of the information about the Universe, and secondly, the furthest and
thus the oldest observable galaxies are actively forming stars.

The near-by blue compact galaxies are the closest approximation of
these early days. They have low metallicities and form stars at, for
their sizes, extraordinarily high rates. Blue compact dwarf (BCD) galaxies,
with sizes of less than 2\,kpc and mostly centrally concentrated star formation
\citep{hunter2006}, are also locally abundant and thus ideal to study
the star-formation processes in detail. Their metallicities range from
half to 1/30$^{th}$ of the solar value \citep{kunth1988}.  This class
of galaxies can be divided further, depending on the type of galactic
body hosting the burst: E-BCDs and I-BCDs, for bursts in smooth dwarf
ellipticals or in irregular bodies, respectively, or by the location
of the burst: ``n'' for a nuclear concentrated burst or ``i'' for
pockets of star formation distributed all over the galactic body. For
example, iE-BCD will stand for a BCD with a smooth elliptical body and
irregularly distributed pockets of star formation \citep{loose1986}.

\begin{table*}
\addtolength{\tabcolsep}{-4pt} 
\caption{Basic characteristics of the galaxies in our sample. The
  columns are as
  follows: name of the galaxy, right ascension and declination 
  in J2000 epoch, heliocentric radial velocity in \kms, distance in
  Mpc, inclination, size of the galaxy in arcsec and in kpc, total \hone\
  mass in solar masses; mass to light ratio in $B$;
  star formation rate inferred  from \hone\ observations; gas
  metallicities, ellipticity ($\epsilon = 1-b/a$), number of galaxies
  in the group. Columns 3,4,5, 6,7,8,9,10,11 are from
  \citet{vanzee2001}; columns 2,12,13 are from the HyperLeda database. }
\begin{tabular}{lcccccccccccccc}
\hline
\hline
Name   & RA DEC  & V$_{Helio}$ & Distance& Inclination &
\multicolumn{2}{c}{D$_{25} \times$ d$_{25}$} & M$_{HI}$ &
M$_{HI}$/L$_B$ & SFR$_{HI}$ &  & $\epsilon$ & N$_{gal}$ \\
        &  J2000   & (\kms)  & (Mpc)     & (deg)   &
        (arcsec) & (kpc) & (10$^8$ M$_\odot$) & (M$_\odot$/L$_\odot$)
        & (M$_\odot$/yr) & 12+log(O/H)     &    &        \\
\hline 
Mk324 & J232632.82+181559.0& 1600 & 24.4 & 38 & 29$\times$23 & 3.4$\times$2.7 &
3.28 & 0.50 & 0.065 & 8.50$\pm$0.20 &0.09 & 8 \\
Mk900 & J212959.64+022451.5& 1155 & 18.0 & 43 & 44$\times$36 & 3.8$\times$3.1 &
1.55 & 0.21 & 0.088 & 8.74$\pm$0.20 & 0.38 & 1 \\
 UM038 & J002751.56+032922.6& 1378 & 20.3 & 40 & 32$\times$24 & 3.1$\times$2.4 &
 2.90 & 0.72 & 0.038 & 8.15$\pm$0.20 &0.12 & 3\\
 UM323 & J012646.56-003845.9& 1915 & 26.7 & 43 & 24$\times$16 & 3.1$\times$2.1 &
 4.23 & 1.03 & 0.111 & 7.70$\pm$0.20 & 0.24 & 10\\
\hline
\label{table:sample}
\end{tabular}
\end{table*}

While star formation in dwarf galaxies (like dwarf irregulars, dIrrs)
is not unusual, it is the intensity  and the concentration of the bursts in the BCDs that
makes them special.  The BCDs are usually more compact than the dIrrs
and experience massive bursts of star formation with high specific
star-formation rates (the star-formation within the scale length, \citealt{hunter2004}).
This, combined with their typical gas content of about
50\,percent of their mass or more \citep{zhao2013}, results in gas
exhaustion times of less than 1\,Gyr \citep{gildepaz2003}. Hence, we
must be witnessing an extraordinary process, possibly transforming
these galaxies into early type dwarfs.

Dwarf galaxies are not efficient at transforming their gas into
stars. The reasons are that, on one side, the gas density is lower in
dwarfs, and on the other, that their shallow gravitational potentials
allow even a few supernova explosions to heat the gas, ceasing further
condensation into stars in almost the full galactic body. These
observationally proven and theoretically backed facts
\citep{bigiel2010,hunter2012,schroyen2013} are in contradiction with
the observed bursts in BCDs. A possible solution to this problem can
be offered by sequential triggering of star-formation
\citep{gerola1980}, cloud impact \citep{gordon1981}, mergers
\citep{bekki2008}, tidal effects \citep{vanzee1998}. Based on the
disturbed H$_\alpha$\ velocity fields and irregular optical
morphologies of the BCDs, many observational studies support that the
burst of star formation is triggered by dwarf galaxy mergers or gas
accretion \citep[e.g.][]{ostlin2004} .

Here, we will try to distinguish between different burst-triggering
mechanisms by comparing the stellar, ionized gas and 
neutral gas kinematical data of a sample of blue compact dwarfs. We
will also study their stellar population properties and try to
anticipate the possible outcome of a BCD when the burst is over. Our
paper is organized as follows: in Sect.\,\ref{sect:data} we will
present the data and describe the analysis tools, in
Sect.\,\ref{sect:results} we will present our results which will be
followed by discussions (Sect.\,\ref{sect:discussions}) and
conclusions (Sect.\,\ref{sect:conclusions}).

\section{Data \& Analyses}
\label{sect:data}

To investigate the resemblance between dwarf ellipticals and BCDs we
selected four E-BCDs from the sample of \cite{vanzee2001}.  The basic
characteristics of these objects are described in
Table\,\ref{table:sample} and their ($B-I$) images are displayed in
Fig.\,\ref{fig:BmI}. Since their bodies have a smooth appearance of an
early type dwarf, one may expect that if the gas is removed/exhausted
and the burst fades they may look like dwarf ellipticals. Their
$B$-band magnitudes (Table\,\ref{table:phot}) are similar to the
luminosities of recent spectroscopical studies of early type dwarfs
\citep[][to name a
few]{derijcke2003,lisker2006,chilingarian2007,koleva2009,toloba2011},
albeit at the low end. Thus, a comparison between different
morphological types is possible.

\subsection{Data}
We were awarded 26 hours of VLT time to perform spectroscopy and
photometry on this sample. In the fall of 2003 we obtained $J$-band
images with the ISAAC and $B$-, $I$-band images and long-slit spectra
using FORS2. Details about the instrumental setup and the exposure times
 are listed in Table\,\ref{tab:setup} and
Table\,\ref{tab:log}.

\subsubsection{Photometry}

The standard data reduction steps for the $B$- and $I$-images were
performed using {\sc
  midas}\footnote{\url{http://www.eso.org/sci/software/esomidas/}}. First
a master bias was subtracted and the images were divided by the
normalized master twilight flatfield obtained a few days from the
observations. The sky was computed using the {\sc fit/flat\_sky}
procedure. This procedure consists of manually selecting clear sky
regions, fitting them with 2 dimensional polynomials (in this case
with degrees of 1,1), averaging this region employing
$\kappa$-$\sigma$ clipping (with $\kappa$ = 2), and finally
subtracting it from the original image. Then, the individual exposures
were median averaged, performing cosmics rejection. Finally, the
images were extinction-corrected and flux-calibrated using the
photometric standard stars observed during the same nights as the
objects, and reduced with the same procedure.

The basic data reduction for the $J$-band images was done with the ESO
pipeline. Background subtraction was done in a 3-step procedure.  In
the first step, the background is subtracted from each individual
image by using a running mean on each pixel. From the background
subtracted images a stacked image is created, which is used to create
an object mask. In the second step the object mask is used on all
individual input images to compute, for each image, a clean background
using a $\kappa$-$\sigma$ clipping procedure. The background subtracted
images were co-added. It turned out that there remained some large
scale pattern background (which we attribute to the
detector). In a third step we computed a `smooth' background by again
applying the object mask to the co-added image and using a high-order
polynomial to fit the remaining background. We used the median zero
point provided by ESO (25.062\,mag) which we cross-checked with our
observations of standard stars.

\begin{table}
  \begin{minipage}{\columnwidth}
    \caption{\label{tab:setup} Setup of observations }
    \begin{tabular}{lcc}
      \hline
      \multicolumn{2}{c}{FORS2 $@$ VLT-U4 } \\
      \hline
      CCDs, MIT/LL mosaic\#   & CCID20-14-5-3 \\
      \# of pixels            & 2 $\times$ 4096$\times$2068 chips \\
      pixel size [$\mu$m$^2$] & 15$\times$15 \\
      image scale [arcsec pix$^{-1}$]  & 0.126      \\
      readout noise [e$^-$ pix$^{-1}$] & 2.7 / 2.9 \\
      gain [ADU (e$^-$)$^{-1}$]        & 0.8 / 0.7  \\
\\
      \multicolumn{2}{c}{Imaging, Program 072.B-0108(B)} \\
\\
Filter & B\_BESS / I\_BESS \\
\\
\multicolumn{2}{c}{Spectroscopy, Program 072.B-0108(A)} \\
\\
grism                      & GRIS\_1028z  \\
slit width [arcsec]        &    0.7   \\
spectral range [\AA]    &  7700 -- 9300    \\
FWHM $\delta\lambda$ [\AA]     &    \\
$\sigma_{instr}$ [km s$^{-1}$] &     \\
dispersion [\AA\,pix$^{-1}$]   &   0.42 \\
      \hline
      \hline
      \multicolumn{2}{c}{ISAAC $@$ VLT-UT3 } \\
      \hline
\multicolumn{2}{c}{Program 072.B-0108(C)}\\
Rockwell Hawaii Hg:Cd:Te arrey& \\
\# of pixels & 1024$\times$1024 \\
pixel size [$\mu$m$^2$] & 18.5$\times$18.5 \\
image scale [arcsec pix$^{-1}$]  & 0.1484      \\
readout noise [e$^-$ pix$^{-1}$] & 11 \\
gain [e$^-$ ADU$^{-1}$]        &4.5  \\
filter & J \\
filter $\lambda_c$  [$\mu$m] & 1.25\\
filter FWHM [$\mu$m] & 0.29\\
\hline
    \end{tabular}
  \end{minipage}
\end{table}

\begin{table}
  \caption{Observations log}
  \begin{tabular}{crrrr}
\hline
    Object & Filter &  Exposure, s & PA, deg & note \\
\hline
    
Mk324  &$B$ &  5$\times$38& - & - \\
&$I$ &  6$\times$27 & - & - \\
&$J$ & 12$\times$30 & - & - \\ 
&GRIS\_1028z & 5$\times$1400 & 77 & minor\\
&GRIS\_1028z &10$\times$1400 & 167 & major\\
Mk900  &$B$ & 5$\times$38 & -& -\\
  &$I$ & 6$\times$27 & -& -\\
  &$J$ &  8$\times$30 & -& -\\
&GRIS\_1028z  & 7$\times$1314 & 178 & minor\\
&GRIS\_1028z&10$\times$1314 & 59 &major\\
Um038   &$B$ & 5$\times$38& -& -\\
   &$I$ & 6$\times$27& -& -\\
   &$J$ &  8$\times$30 & -& -\\
&GRIS\_1028z& 6$\times$1400 & 178 &minor\\
&GRIS\_1028z&10$\times$1400 & 88 &major\\
Um323   &$B$ & 5$\times$38& -& -\\
   &$I$ & 6$\times$27& -& -\\
   &$J$ & 8$\times$30 & -& -\\
&GRIS\_1028z& 6$\times$1400 & 174 &minor\\
&GRIS\_1028z&13$\times$1400 & 84 &major\\

\hline
  \end{tabular}
  \label{tab:log}
\end{table}

\begin{figure*}
\includegraphics[width=0.24\textwidth]{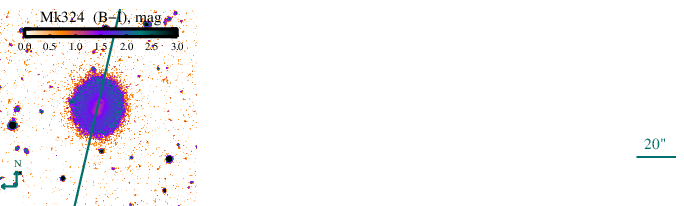}
\includegraphics[width=0.24\textwidth]{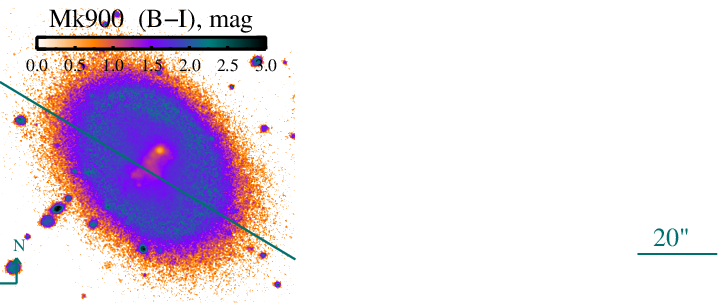}
\includegraphics[width=0.24\textwidth]{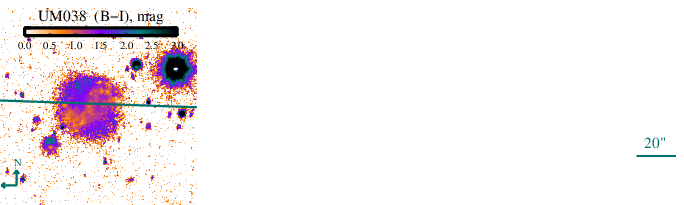}
\includegraphics[width=0.24\textwidth]{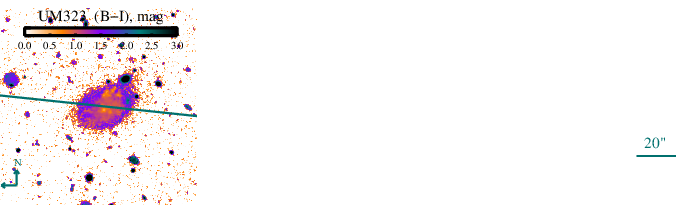}
\caption { From left to right Mk324, Mk900, UM038, UM323 colour ($B-I$)
  maps. We overplot the major axis positions of the slit, coinciding
  with the angle of maximum \hone\ rotation \citep{vanzee2001}.}
\label{fig:BmI}
\end{figure*}

\subsubsection{Spectroscopy}

The spectroscopic observations were performed in the wavelength range
7700--9300\,\AA\ with a reciprocal instrumental resolution
($R=\lambda/\delta\lambda$) of 5000. The slit aperture was set to
0.7\arcsec\ and the seeing varied between 0.51-1.44\arcsec. We
performed a standard data reduction, consisting of bias subtraction,
flat-fielding and wavelength calibration with the {\sc
  midas}
software package. The precision of the wavelength calibrations was
checked with \ulyss, using the HD201381 (G8III) calibration star
observed during our run against the CFLIB interpolator
\citep{wu2011}. This interpolator is a function,
which returns a stellar spectrum for a given temperature, gravity and
metallicity. We found that the precision of the wavelength calibration
is 2\,\kms (1/10$^{th}$ of the original pixel size), while the
instrumental velocity dispersion is around 27$\pm$2\kms\ (corrected
for the model resolution, see Sect.\,\ref{subsec:spectroscopy} for more details),
across the full wavelength range. We then used the dispersion relation
to derive and subtract a two-dimensional model for the sky background
prior to the wavelength calibration, using the method proposed by
\citet{kelson2003}. Given the very strong and, fortunately, also
curved sky emission lines in this wavelength region, this method
provided us with significantly smaller sky residuals than the standard
sky subtraction algorithms implemented in e.g. {\sc midas}.

\subsection{Analyses}
\subsubsection{Photometry}
\label{subsec:phot} 
After a careful removal of the sky background, we interpolate over all
stars superposed onto the galaxy images. We use a custom computer code
to derive the photometric characteristics of the galaxies in the
sample. Given an estimate for the galaxy center, the code fits the
galaxy's isophotes with ellipses whose center, axis ratio, and
position angle are free parameters \citep{derijcke2003}. The
deviations of the isophotes from a pure elliptic shape were quantified
by expanding the intensity variation along an isophotal ellipse in a
fourth order Fourier series. This allows for an accurate determination
of the center of the outer isophotes which can be used as an estimate
for the "true" center of the galaxy. Around this center, we used
circular isophotes to estimate the total luminosity and the circular
half-light radius of each galaxy. Given the irregular nature of these
galaxies, such a model-independent way of representing their
photometric properties seems appropriate.

The $(B-I)$ images (Fig.\,\ref{fig:BmI}) were produced in the
following way: prior to the subtraction, we first trimmed the images
to 27.2 mag, such that all the values above 27.2 were set to 27.2. The
later value is the photometric zero point of Mk324 $B$-band image.

\subsubsection{Spectroscopy}
\label{subsec:spectroscopy}
We derive the radial distribution of the kinematics and stellar
population properties of our sample by means of full spectrum
fitting. We use \ulyss\footnote{\url{http://ulyss.univ-lyon1.fr}}
\citep{ulyss}, a flexible, freely available package, which minimizes
the difference between a target spectrum (observations) and
interpolated models (templates), using the following expression:

\begin{align}
{\rm{ Obs}(x)} = P_{n}(x) \times \bigg(
    &LOSVD(v_{sys},\sigma) \nonumber \\
    &\hspace*{-3em}\otimes \sum_{i=0}^{i=m} W_i \, CMP_i(a_1, a_2, ...,x) \bigg)   + A_{l}(x).
\label{eqn:main}
\end{align}

The observations ($Obs(x)$) and the templates ($CMP_i(a_1, a_2,
...,x)$) are logarithmically rebinned to the same constant step in
velocity prior to the minimization, $x = ln(\lambda)$, where $\lambda$
is the wavelength in Angstroms. The multiplicative polynomial $P_n(x)$
(of a degree $n$) absorbs differences between the model and the
observations, resulting from a bad (absent) flux calibration and poor
knowledge of the extinction law, while the additive polynomial
($A_l(x)$, of a degree $l$) can be used when a hot (featureless)
component is expected in the spectrum. We tested an inclusion of
$A_l(x)$ in our fits and we concluded that such a component is absent,
confirming the adequacy of the sky subtraction. The line-of-sight
velocity distribution (LOSVD) is characterized by the object's radial
velocity or the shift of the lines ($v$) and their broadening
($\sigma$). In principle, higher orders of the LOSVD can be used, but
in our case the quality of the spectra is not sufficient to attempt
fitting of the lines' kurtosis (for symmetric departures from a Gaussian)
and skewness (departures from symmetry).

Here we use four different representations of this
equation.
\begin{itemize}
\item To derive the instrumental broadening 
  (see further) we minimize a spectrum of a star against an 
  interpolator of an empirical spectral stellar library ({\sc TGM} component
  in \ulyss) as in
  \cite{wu2011,prugniel2011,koleva2012}. In this way we determine the best fitted
  temperature, gravity, and metallicity ($a_1,a_2,a_3$), together
  with the instrumental broadening. We assume
  that the physical broadening in the stellar lines is negligible, thus
  the derived dispersion is a result only of the instrumental
  dispersion minus the template broadening (squared). Hence,
  Eq.\ref{eqn:main} degenerates to:
\begin{align}
{\rm{ Obs}(x)} = P_{n}(x) \times
    LOSVD(v_{sys},\sigma) \otimes TGM(a_1, a_2, a_3,x) 
\label{eqn:inst_disp}
\end{align}

\item To derive the instrumental dispersion of the models
  (see further) we fit only the LOSVD between two
  stars. Then we have:
\begin{align}
{\rm{ Star_1}(x)} = LOSVD(v_{sys},\sigma) \otimes Star_2(x) 
\label{eqn:temp_disp}
\end{align}

\item When we use \ulyss\ to get the kinematical and stellar
  population properties of our galaxies, $CMP$ represents a single stellar
  population (SSP) from an interpolated grid of population synthesis
  models. Here, the SSP is a function of the age ($a_1$) and
  the metallicity ($a_2$) of the model and we use:
\begin{align}
{\rm{ Obs}(x)} = P_{n}(x) \times 
    &LOSVD(v_{sys},\sigma) \otimes SSP(a_1, a_2,x).
\end{align}

\item To derive the gas kinematics, the components ($CMP$) are
  Gaussians ($GAUSS$), whose individual broadenings ($\sigma^\star$) and
  positions ($x$) are adjusted
  during the fit, or:
\begin{align}
{\rm{ Obs}(x)} = P_{n}(x) \times 
   \sum_{i=0}^{i=m} W_i \, GAUSS_i(\sigma^\star_i,x) .
\label{eqn:main}
\end{align}

\end{itemize}

The CaT (calcium triplet) region and the CaT index in particular, are
usually used as a tracer of the dynamical state of the old stellar
population in early type systems \citep{delisle1992}. However, for
relatively young and metal-poor systems, with SSP-equivalent ages
below $\sim$3\,Gyr and metallicities below -0.5\,dex, this region can
provide valuable information about the underlying stellar population
\citep{vazdekis2003}.

However, the constraints on the ages and metallicities we can obtain
  from observations do not necessarily comply
with theoretical expectations. The latter are derived from very high signal-to-noise
stellar population models and do not include the effects of template mismatch.
To investigate to what degree the region we fit can constrain the ages
and metallicities of galaxies, we extracted one arcsec wide, 1D
spectra at the luminosity peaks of the galaxies and we constructed
$\chi^2$-maps (See Fig.\,\ref{fig:chimaps}).  The signal-to-noise ratio
(S/N) at the luminosity peak is around 25. We find that, in most
cases, we cannot recover ages higher than 1.2 to 1.5\,Gyr and the
precision of the metallicity is around 0.2\,dex. Even at this
relatively high S/N the age resolution we get is lower than predicted
from theory. This is expected not only because of the intrinsic
imperfectness of the data, but also because of the template-mismatch
since our knowledge of the stellar evolution in this spectral region
is less advanced than in the optical.

\begin{figure}
\includegraphics[width=0.49\textwidth]{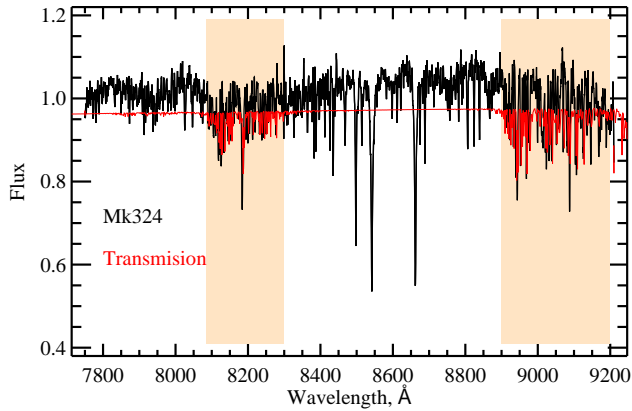}
\caption{\label{fig:atm} Relative atmospheric transmission (with red) in the CaT region. As 
  a reference we plot the central normalized spectrum of Mk324 (with
  black). The shaded areas are the spectral regions wich were masked. }
\end{figure}

Prior to the fitting we exclude the regions which are strongly
affected by the telluric absorption. We
computed the atmospheric transmission using the ESO web tool for sky
modeling\footnote{\url{http://www.eso.org/observing/etc/bin/gen/form?INS.MODE=swspectr+INS.NAME=SKYCALC}}
for our resolution, wavelength coverage and for average sky
conditions. The result of the transmission curve can be seen on
Fig.\,\ref{fig:atm}.  From this figure it is clear that the regions
between $\lambda \in [8085-8300]$\,\AA\ and from
$\lambda > 8900$\,\AA\ onwards are strongly affected by tellurics and
we excluded them from our fits. However, we kept the region around
\sthree at 9068.6\,\AA, since this emission line is the strongest in our spectra.

Throughout this paper we use the Pegase.HR population synthesis models
\citep{pegase.hr} in combination with the CFLIB empirical spectral
library \citep{valdes2004,wu2011} as templates. Our choice of models
was driven by the fact that CFLIB contains a reasonable number of
stars (1273) covering well the primary atmospheric parameters
(effective temperature, gravity and metallicity), and has a high
enough spectral resolution (R$\sim$5000) to allow us to determine the
stellar population parameters and the velocity dispersions of our
galaxies. To be able to compute the physical dispersion of our
galaxies we need to add the template dispersion and subtract the
instrumental broadening, quadratically. In \cite{wu2011} the precise
spectral resolution of CFLIB was assessed, but only for the optical
spectral range ($4500$\,\AA\ $< \lambda< 6500$\,\AA).  Here, we aim to
derive the spectral resolution in the CaT region. First, we selected a
solar-analog star\footnote{A star with atmospheric parameters similar
  to the Sun.} from the CFLIB library - HD010307. We fitted
(using Eq.\,\ref{eqn:temp_disp}) its spectrum against an observation of the
Sun, from the HyperLeda database with R=10000
($\sigma\approx13$\,\kms) in the region 8400-8900\AA. This region is
free from telluric lines and has prominent spectral features. We find
a velocity dispersion of $\sigma\approx13$\,\kms, which translates to
a resolution of $\sigma\approx18$\,\kms, when corrected for the
template broadening. However, the produced final stellar population
models can have slightly higher broadening since the individual
stellar spectra are interpolated and added. To check this, we extract
a solar analogue from the CFLIB interpolator and again fit it against
the same observations of the Sun. We derive a total broadening of
19\,\kms, consistent with the value derived from individual star
  fitting. We use this value throughout the paper.
 
\section{Results}
\label{sect:results}

The results of our photometric analyses are presented in
Table\,\ref{table:phot}.  In young populations, the $B$-band (around
440\,nm) is typically dominated by A-type main sequence stars (hot
stars), while the $I$- and $J$-bands are dominated (after the first
$\sim$0.5\,Gyr of the star formation) by (cold) red-giant branch stars. The ($B-I$) colour
maps are sensitive to young ages but also to metallicity and
gravity. In other words, the ($B-I$) colour images can show us the locus
of the young population, but can also include effects from the
dwarf-to-giant ratio and varying metallicity.  Regions where
$(B-I)<0.4$ are dominated by horizontal branch stars and young stars
on the main sequence. Hence, these values on the colour images indicate
possible young populations with ages less than 1\,Gyr.

We do not discuss the ($B-J$) images since they are much noisier and
do not bring new information. The ($I-J$) images are more sensitive to
the metallicity of the RGB. We compute an average ($I-J$) colour of 4.5
with little variations around this value. However, we have spectra
which are more sensitive to smaller metallicity changes. 

The spectroscopic results from our major and minor axis analyses are
presented in Fig.\,\ref{fig:Mrk324_prop} --
Fig.\,\ref{fig:UM038_prop}. The mean values of the stellar parameters
(velocity, velocity dispersion, ages and metallicities) are reported
in Table\,\ref{table:spec}.  The data were binned to a S/N of 20.  The
size of the pixels is around 0.25\,arcsec, but the seeing during our
observations was around 1\,arcsec.  Hence, results below 1\,arcsec are
not independent. 

Before discussing the kinematics of the stars we should remember that
dwarf galaxies have close to exponential surface brightness profiles
and (if isotropic) are expected to reach their maximum velocity at
about 2$\times$ their effective radius \citep{prugniel2003}. Thus,
their reported stellar maximum velocities are only lower limits for the
maximum rotation.

A second caveat comes from the fact that since these galaxies are
star-forming, it is somewhat tricky to identify the center of the main
stellar body. Our galaxies experience mostly off-centered star
formation, which seems to have a large influence on their kinematical
profiles. To identify the centre of the stellar body we fitted a
Gaussian to the luminosity profile through the slit, excluding the
main star-formation regions, identified with the help of the ($B-I$)
images (Fig.\,\ref{fig:BmI}). 

Finally, the last warning is about the stellar population properties of
the galaxies. Though we could measure SSP-equivalent ages and
metallicities, they are uncertain especially for ages more than a few
Gyr, where the CaT region does not change significantly with time
anymore. In other words,  the stellar populations with measured ages above 1\,Gyr can be
  anywhere between 1 and 13\, Gyr old.

We find that our galaxies have ages between 1 and 4\,Gyr and metallicities between
$-$1.6 and $-$0.4\,dex.  Their stellar population profiles are rather
flat up to one effective radius, but again, as for the kinematics,
these profiles are not easily described with a simple linear law. From
the $\chi^2$-maps (Fig.\,\ref{fig:chimaps}), we see that the
determination of the age is not straightforward  and the parameter
space is basically flat between 1 and 3\,Gyr.  Due to the
age-metallicity degeneracy \citep{worthey1999}, also the determination
of metallicity is insecure within $\pm$0.2\,dex. Thus gradients
smaller than this value will be difficult to detect.  

The velocity dispersions of the four galaxies are roughly between 30
and 40\,\kms.  From our 8 profiles of 4 galaxies we do not detect
significant rotation, moreover their velocity fields are significantly
disturbed. Hence, the v/$\sigma$ stellar ratio is close to 1. However,
the star formation region (corresponding to the flux peaks) sometimes
seems to have rotation, and in general disturbs the global
galactic kinematics.

\subsection{Mk\,324}
A good example in this respect is Mk\,324
(Fig.\,\ref{fig:Mrk324_prop}, left panel), where the bright,
star-formation region, between $-$4 and $-$1\,arcsec along the major
axis seems to be almost completely kinematically decoupled from the
rest of the galaxy -- it has lower velocity dispersion and even a
solid body rotation profile. The other, less prominent star-formation region
around $+$2\,arcsec also has a lower velocity dispersion.  From the
colour maps (Fig.\,\ref{fig:BmI}) we do not see a significant young
population. \cite{martinez2007} studied the \halpha\ emission
of this galaxy using Fabry-Perot observations. Assuming an instantaneous
burst, the authors concluded that 95\,per cent of the \halpha\ emission
comes from 4\,Myr old stars. To check our data sensitivity to such
young populations we performed several simulations, mixing different
mass fractions of a 10\,Myr and a 8\,Gyr old population, including
all observational effects. Based on these
tests, we conclude that
the mass fraction of young stars in Mk324 should be less than 1\,per cent.

Moreover, it is worth noticing that the individual band
images (see Sect.\,\ref{app:photometry}) are showing a pronounced burst,
even visible in the $J$-band image (Fig.\,\ref{fig:imaJ}).  From our
spectroscopic analysis we see that the regions of star formation seem
to have significantly higher metallicity, $-0.4$ dex versus the
$-1.4$\,dex of the galactic body. Their SSP-equivalent ages are also
older (more than 2-3\,Gyr), confirming what we see on the images.  We
can conclude that we have a relatively long lived burst, kinematically
decoupled from the rest of the galaxy, which is massive enough to keep
the produced metals and to self-enrich.  Mk324 is part of a group with
8 members. The closest companion is PGC5065954 (with $i$-band SDSS
apparent magnitude of 16.7\,mag), NE, at around
2\,Mpc. The orientation of the burst region is perpendicular to this
direction. Thus, this star formation activity is unlikely sustained
from constant companion disturbance.

\subsection{Mk\,900}
According to the HyperLeda grouping algorithm, MK900 (NGC7077) has no
known companions.  Mk900 is the other BCD from our sample with 
central star formation.  Looking into detail at the ($B-I$) image
(Fig.\,\ref{fig:BmI}) we see that the peak of the luminosity of the
major axis profile (Fig.\ref{fig:Mrk900_prop}) is actually a region of
recent star formation, which is a bit shifted with respect to the centre of the outer
isophotes. This star-formation region has lower velocity dispersion and
rotates at about 5\,\kms, it has also younger ages (less than
1\,Gyr) and higher metallicity ($\sim-$0.7\,dex) than the rest of the
galaxy, which has a mean age of around 1.7\,Gyr and metallicty around
$-$1\,dex. The $J$-band image is much smoother than in the case of
Mk\,324, but we can see peaks of luminosity in this region.
  The central disk rotation is also visible in the minor
axis profile (Fig.\,\ref{fig:Mrk900_prop}), as well as the disturbance
in the velocity and sigma at $-$4\,arcsec, where another knot of star
formation is observed. This knot seems to have somewhat lower
metallicity ($-$1.4\,dex) than the galactic body.

\subsection{UM038}
For UM038, it seems that the star-formation follows a twisted
structure, almost like spiral arms. These structures are most probably
produced by differential rotation, which is also noticeable on the
\hone\ maps (see Fig.\,\ref{fig:gas_stars}), but only outside
10-20\,arcsec (inside this radius the gas and stars rotate like a
solid body, see Section\,\ref{subsec:star_gas}).  Spiral-like
structures are observed in rotating simulated dwarf galaxies, but they
are not stable \citep{schroyen2013}. UM038 is part of a small group
together with another 2 galaxies. They are both to the east,
north-east of UM038 and at a projected distance of more than 0.2\,Mpc.
The major axis of UM038 (Fig.\,\ref{fig:UM038_prop}) displays, again,
an off-centered peak of luminosity associated with disklike rotation
and a drop in velocity dispersion. The same structure has slightly
higher metallicities ($[Fe/H]=-1.0$\,dex) and younger ages (below
1\,Gyr), but this change is not significant. The images in the three
bands show the star-formation structure, which seems to become
smoother with increasing wavelength.  Disk rotation is also visible in
the minor axis data (Fig.\,\ref{fig:UM038_prop}). The velocity profile
indicates an offset kinematic center with respect to the slit
position, or the luminosity center of the galaxy. This is deduced from
the negative velocities going up until the center and then decreasing
again. If the minor axis slit was aligned with the center of the
rotating disk we would have measured zero rotation, but in this case
we have a '$\Lambda$'-like velocity profile. This is at odds with the
ionized gas rotation (see further), which is an indication of
a misalignment of the stellar and gas rotation axes.  The peak of
luminosity is at about $+$0.5\,arcsec and this region has slightly
lower age and higher metallicity, the star formation there probably
appeared at the same time as at the less prominent peak on the
negative side of the galaxy, which has basically the same age and
metallicity.  From Fig.\,\ref{fig:UM038_prop} we can see that the
maximum velocity we observe in this galaxy is about 10\,\kms, while its
velocity dispersion is around 30\,\kms. Hence, its $v/\sigma \sim
0.33$.  Its lower limit of the age is 1.5\,Gyr and its mean
metallicity is $\sim -1.2$\,dex.

\subsection{UM323}
The star-formation pockets (hottest parts) of UM323 are forming a
crown-like structure which is shifted to the NW with respect to the
center of the outer isophotes of the galaxy. The red object at the
north, north-west end of UM323, is a galaxy at a redshift of
0.484\footnote{\url{http://skyserver.sdss3.org/dr9/en/tools/explore/obj.asp?id=1237663783133118504}}.
The body of UM323 is very regular with a $(B-I)$ colour of 1.5.  The most
luminous region along the UM323 major axis
(Fig.\,\ref{fig:UM323_prop}) displays a more turbulent character with
a slight increase in velocity dispersion. The ages and metallicities are
spread around 1.1\,Gyr and $-$1.6\,dex. Its minor axis
(Fig.\,\ref{fig:UM323_prop}) peak of luminosity seems to be shifted
from the center with about 3-4\,arcsec. The star-forming knot has a velocity
difference with the main part of the galaxy of $\sim
-$15\,\kms. Nothing spectacular happens with the ages and metallicities,
but we have only 2 points of the main galaxy to compare to.

UM323 is
part of a bigger group of galaxies (45 in total). The closest
neighboring galaxies are SDSSJ012838.55-004030.8 (SE) and
RGK2003]J012616.22-002 (W) with the same radial velocities and at
projected distances of more than 0.7\,Mpc. In principle these could have
interacted with UM323.

\begin{table*}
\caption{Photometric properties of our sample. The first column is the
name of the galaxy, followed by the apparent magnitude in $B$-band,
absolute $B$-band luminosity computed using the distances in
Table\,\ref{table:sample}, the effective radius of the $B$-band images in
arcsec and kpc. The same properties are reported for the
$I$-band images.}
\begin{tabular}{lcccccccc}
\hline
\hline
name & $B_t$ & M$_B$ & \multicolumn{2}{c}{\reff($B$) }& $I_t$ &
M$_I$ & \multicolumn{2}{c}{\reff($I$) }\\
& (mag) & (mag) & (arcsec) & (kpc) &(mag) & (mag) & (arcsec) & (kpc)\\
\hline
\input{phot_prop.tex}
\hline
\label{table:phot}
\end{tabular}
\end{table*}

\begin{table*}
  \caption{Spectroscopic properties of our sample. The first column is the
    name of the galaxy, followed by the maximum velocity corrected for asymmetric
    drift in one effective radius, then the maximum velocity
    derived by fitting an atan profile to the direct stellar measurements in \kms, the mean
    values of the velocity dispersion in \kms, the age and
    metallicities, the stellar mass-to-light ratio computed from the
    SSP-equivalent characteristics and finally the corrected
    for inclination (v/$\sigma$)$_\star = $v$_{max} / <\sigma> / \sqrt{\epsilon/(1-\epsilon)}$.}
\begin{tabular}{lccccccc}
\hline
\hline
name & v$^c_{max}$ & v$_{max}$ & $\langle \sigma \rangle$ & $\langle$Age$\rangle$& $\langle$[Fe/H]$\rangle$
& M$_\star$/L$_I$ & (v/$\sigma$)$_\star$\\ 
& (\kms) & (\kms)  & (\kms) & (Gyr) & (dex) & (M$_\odot$/L$_\odot$) &  \\
\hline
\input{table_spec.tex}
\hline
\end{tabular}
\label{table:spec}
\end{table*}

\begin{figure*}
\includegraphics[width=0.49\textwidth]{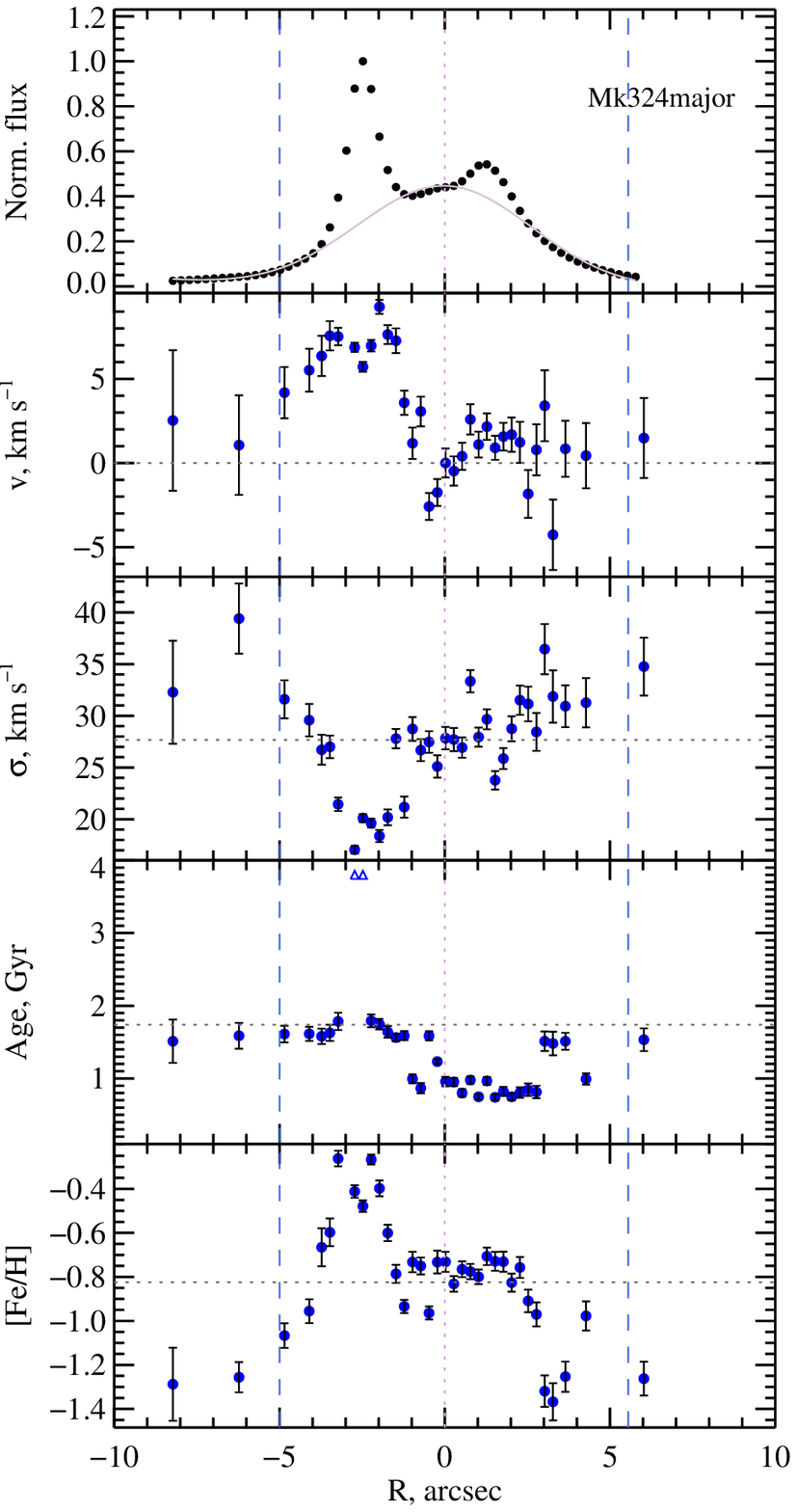}
\includegraphics[width=0.49\textwidth]{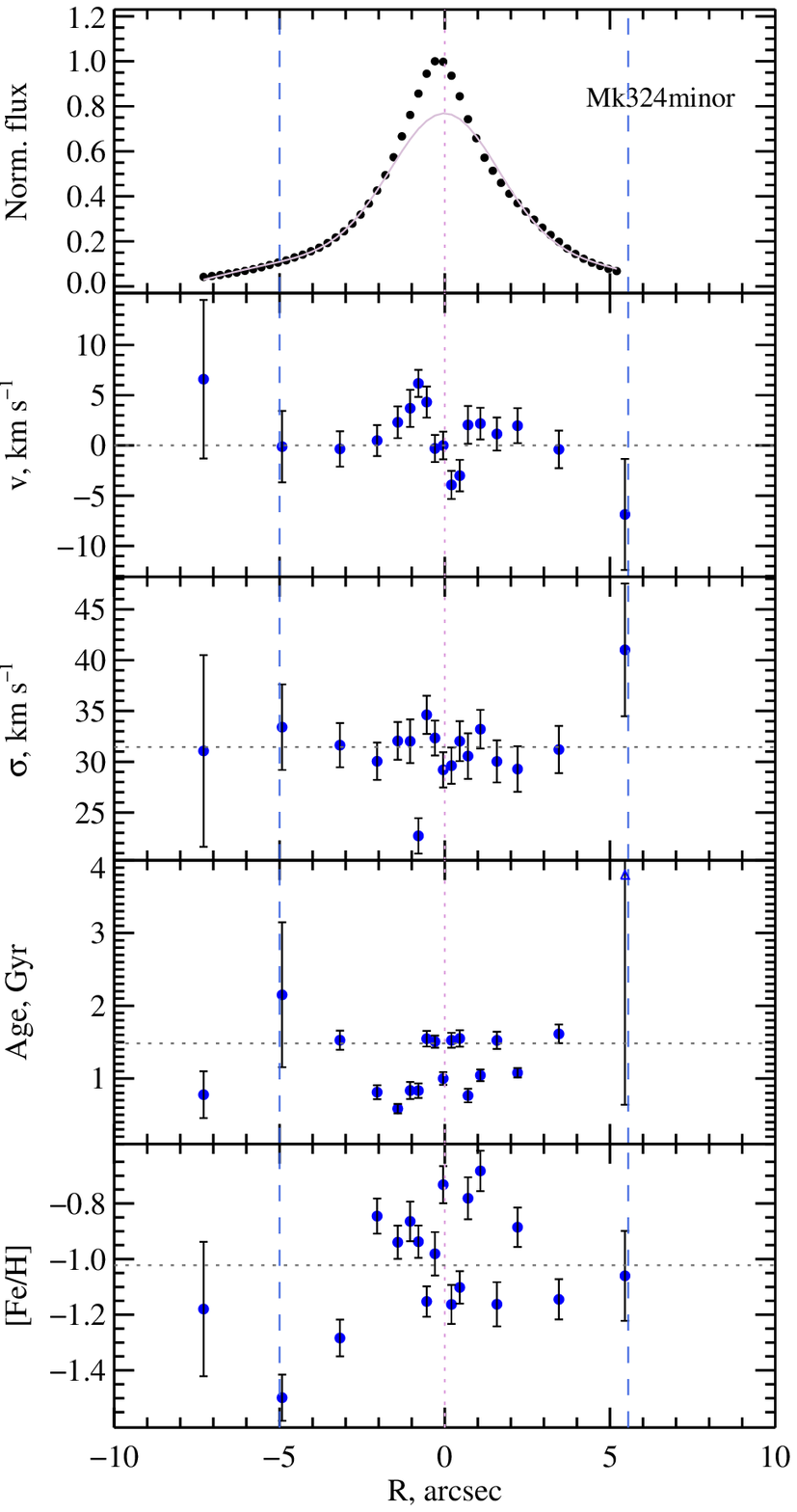}
\caption { Radial profiles of the major and minor axes of Mk900.  We
  did not take the maximum of the flux at the center of the galaxy,
  since it is only indicating the position of the most recent star
  formation event. Instead we did a Gaussian fit (thin purple line
  in the uppermost panel) of the outer parts of the galaxies,
  excluding the star formation regions.  We set the center of the
  galaxy and the center of the Gaussian.  The orientation of the
  major axes can be seen in Fig.\,\protect{\ref{fig:BmI}}, the minor
  axis is perpendicular to it. The upper most panels show the
  luminosity profile within the slit, normalized to its maximum
  point. Going down, the panels are as follows: rest frame velocity in \kms,
  velocity dispersion in \kms, age in Gyr and metallicity in dex. With
  dashed blue lines we show the effective radii in $I$-band derived in
  Sect.\,\protect{\ref{subsec:phot}}. The zero velocity and the
  mean values of the other parameters are plotted with gray dotted
  lines. 
}
\label{fig:Mrk324_prop}
\end{figure*}

\begin{figure*}
\includegraphics[width=0.49\textwidth]{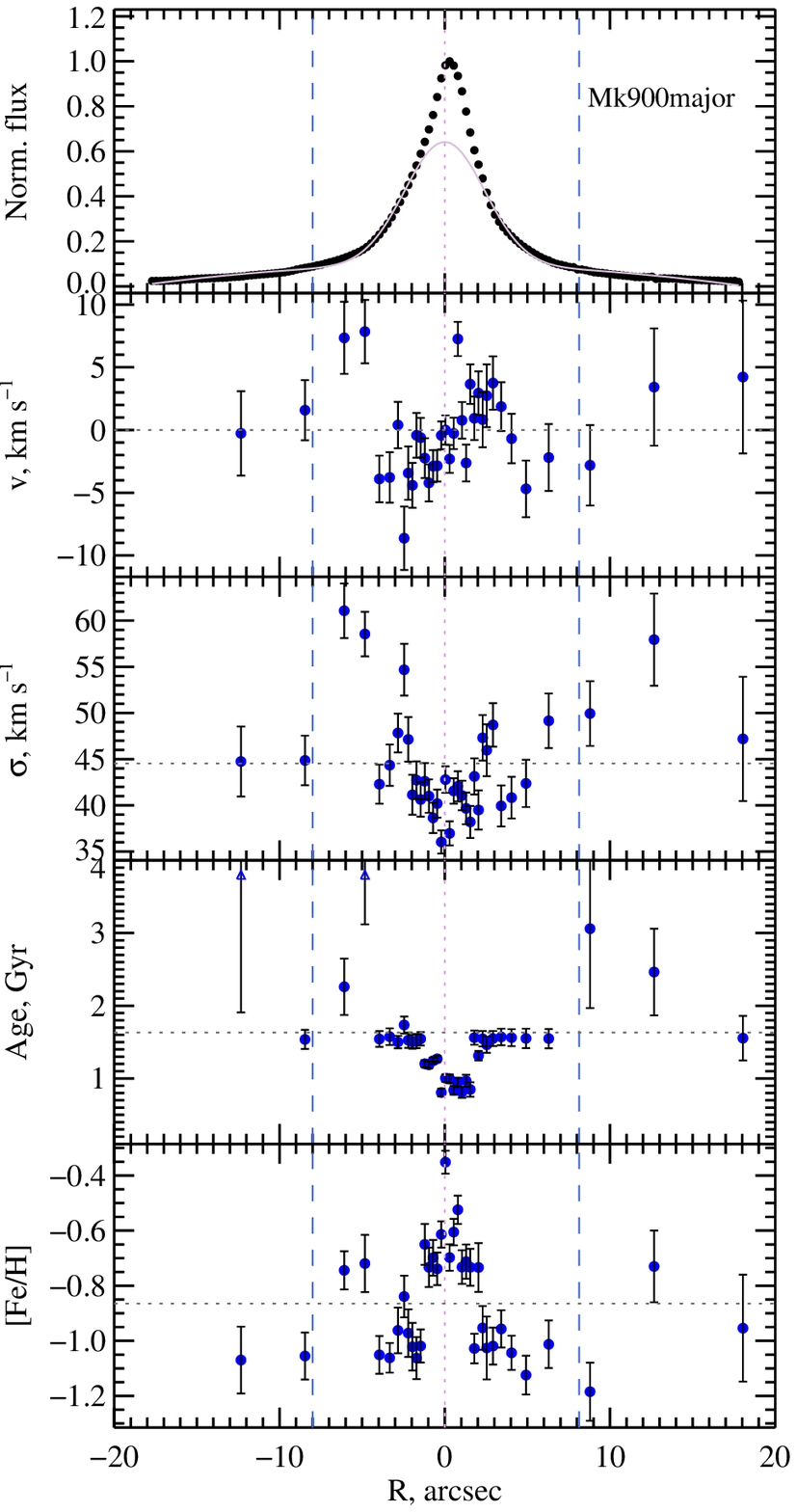}
\includegraphics[width=0.49\textwidth]{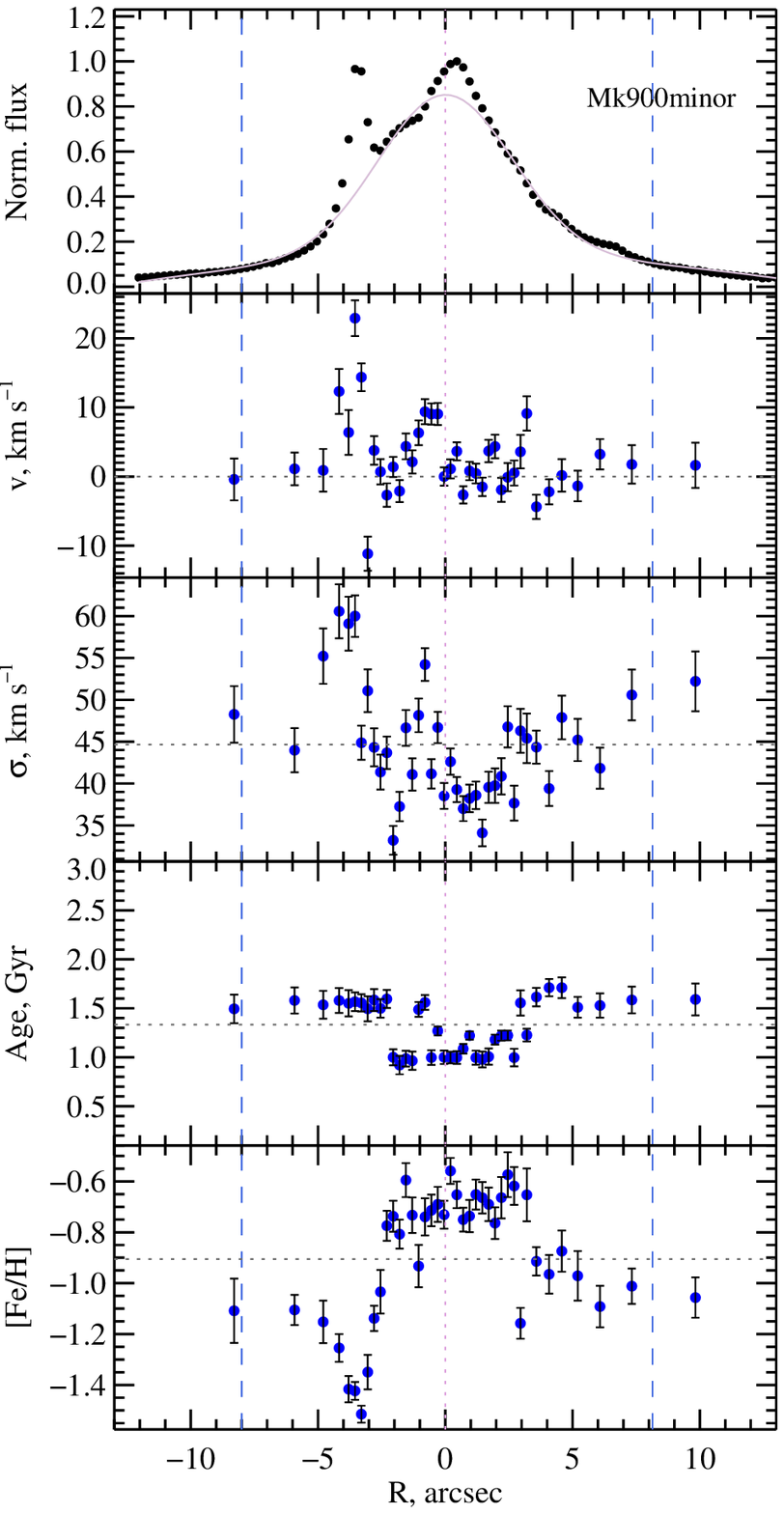}
\caption { Radial profiles of the major and minor axes of
  UM038. Annotations are the same as in Fig.\,\ref{fig:Mrk324_prop}.}
\label{fig:Mrk900_prop}
\end{figure*}

\begin{figure*}
\includegraphics[width=0.49\textwidth]{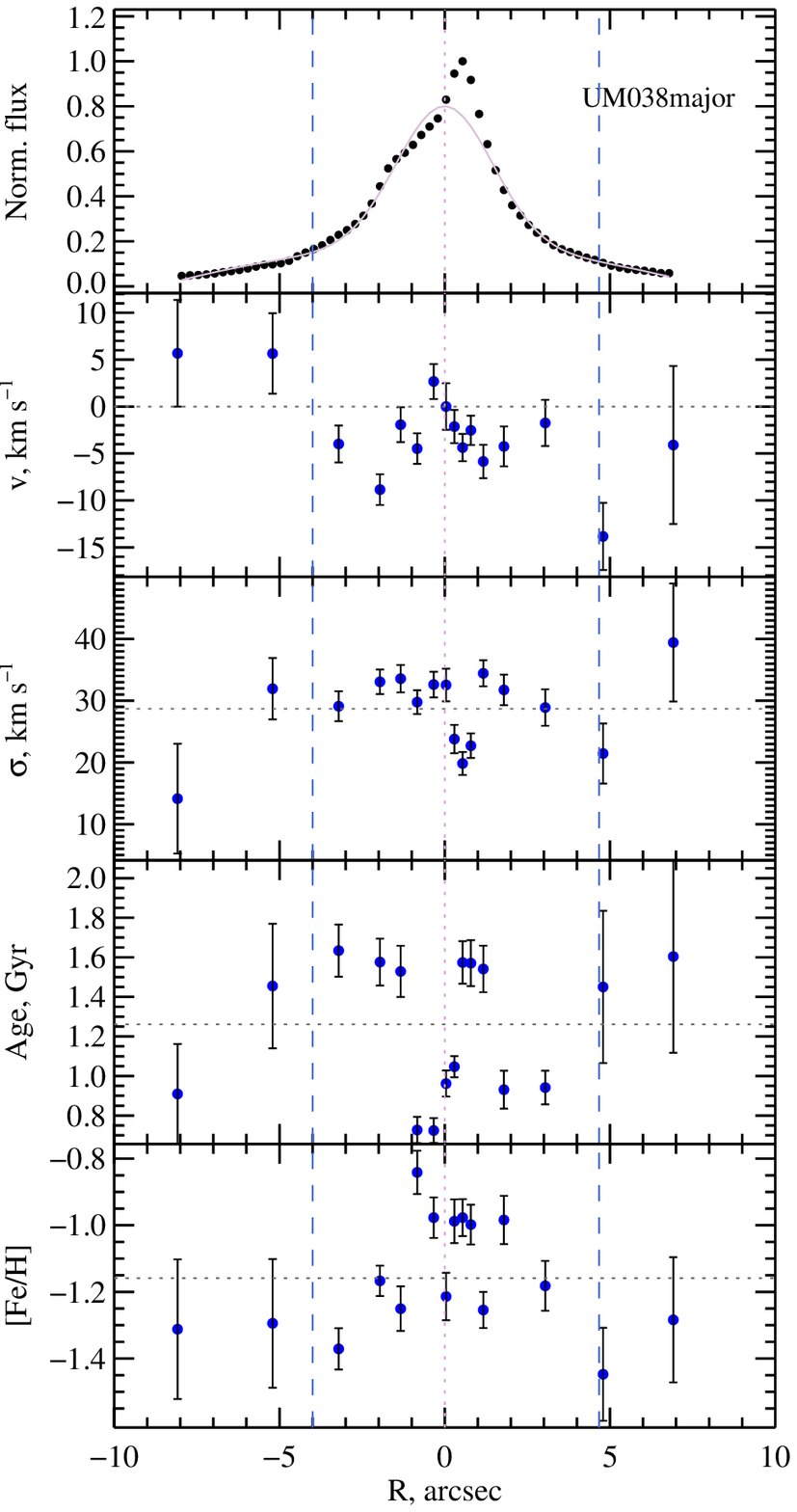}
\includegraphics[width=0.49\textwidth]{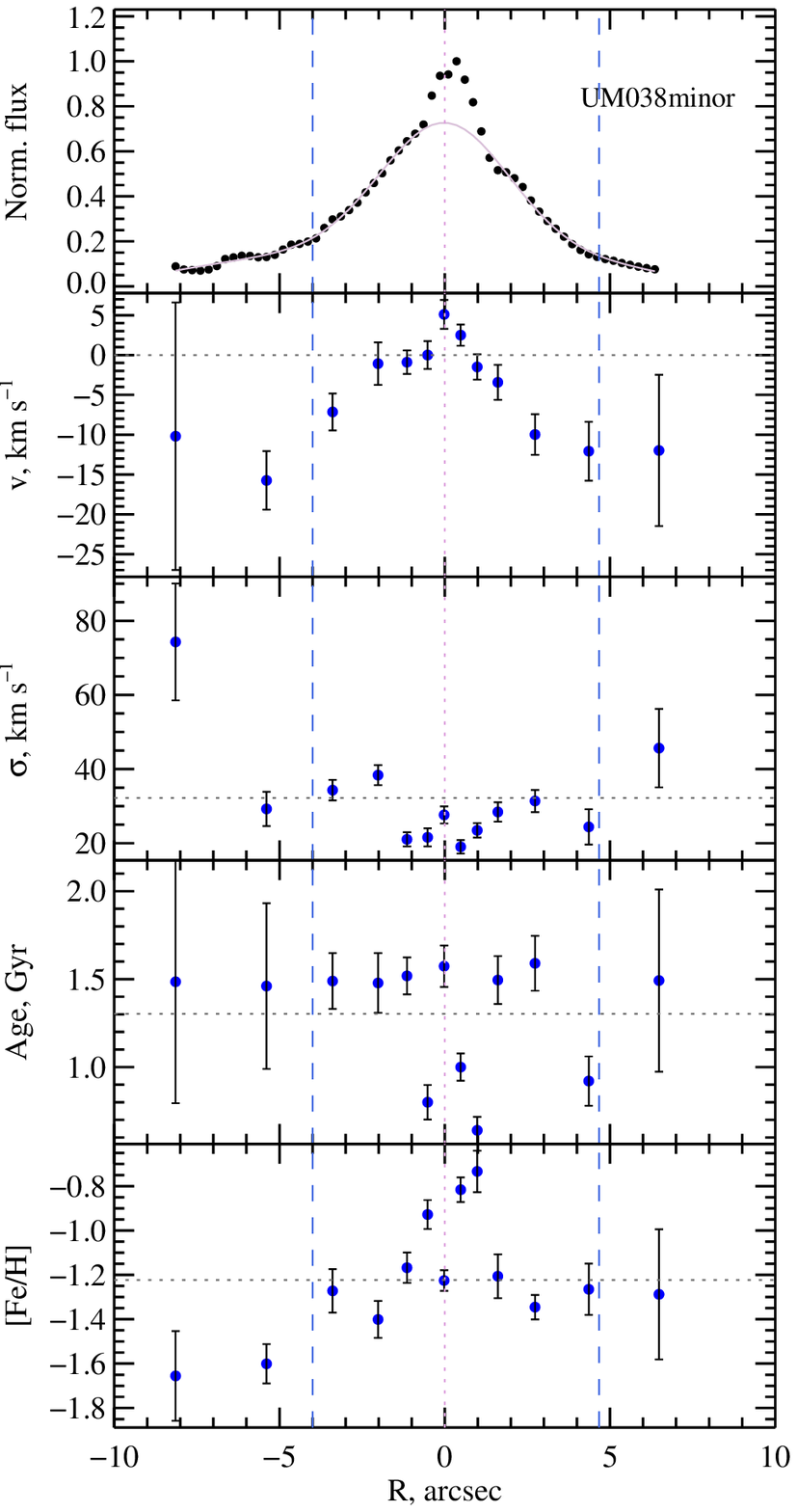}
\caption { Radial profiles of the major and minor axes of
  UM038. Annotations are the same as in Fig.\,\ref{fig:Mrk324_prop}.
}
\label{fig:UM038_prop}
\end{figure*}
\begin{figure*}
\includegraphics[width=0.49\textwidth]{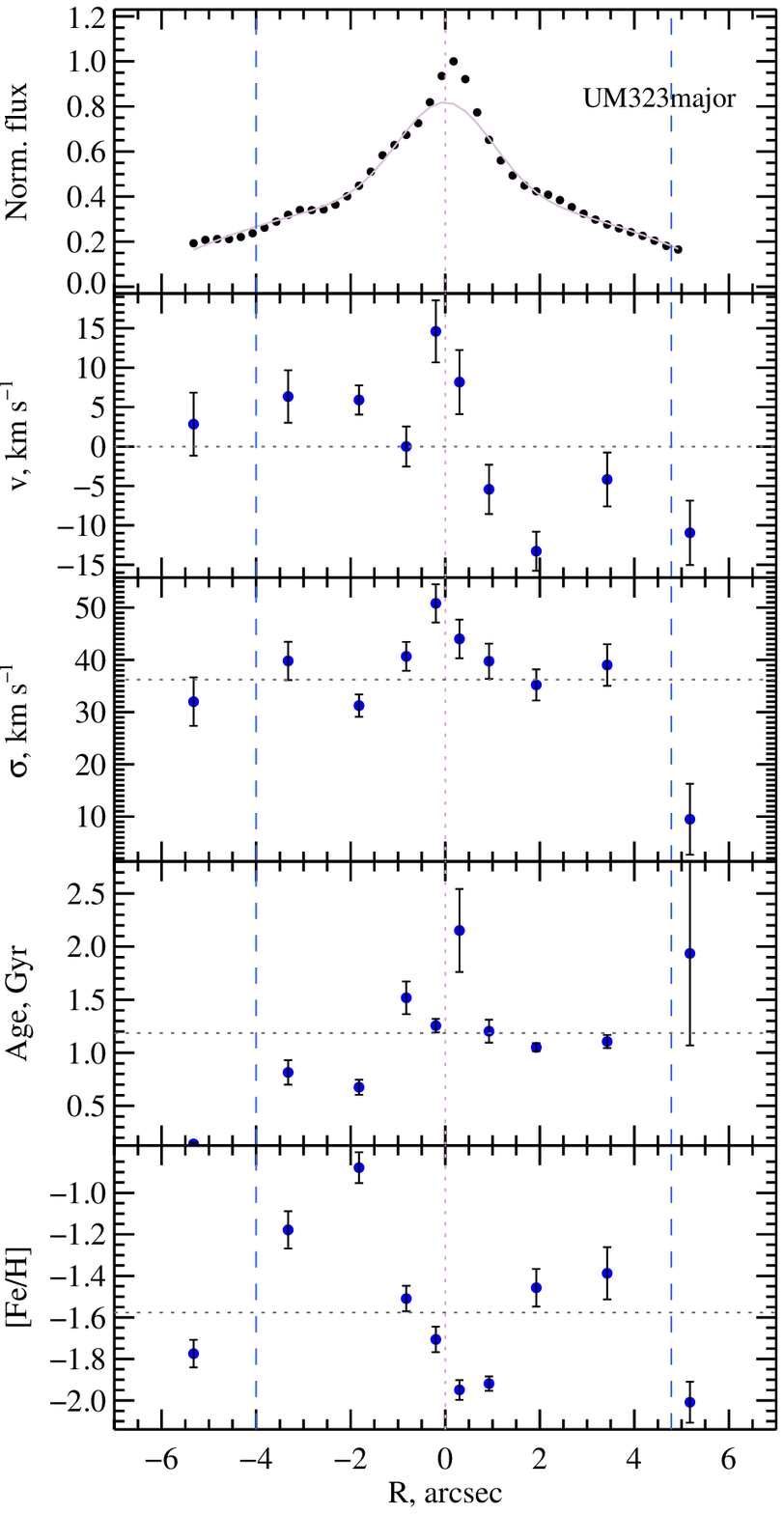}
\includegraphics[width=0.49\textwidth]{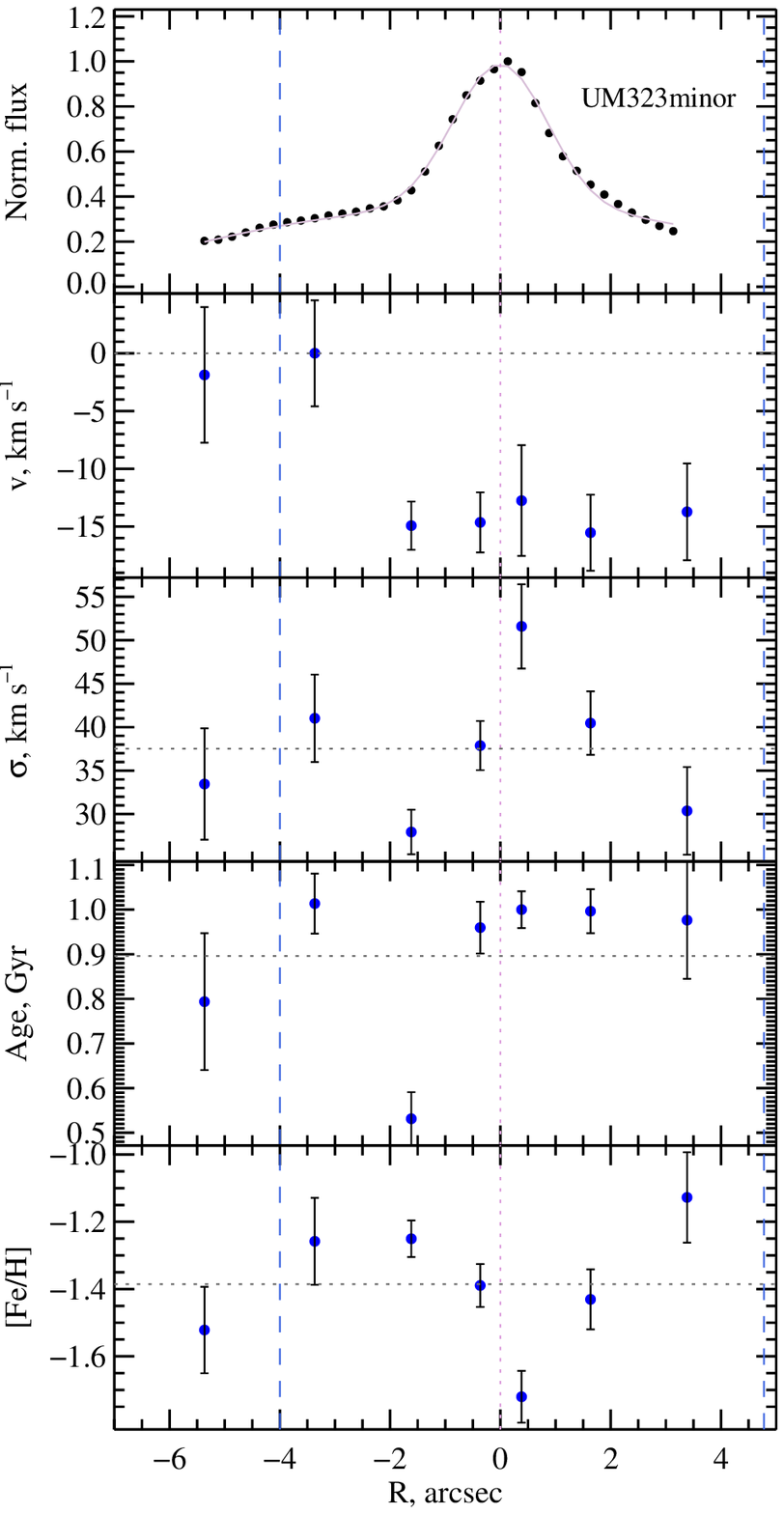}
\caption { Radial profiles of the major and minor axes of
  UM323. Annotations are the same as in Fig.\,\ref{fig:Mrk324_prop}.
}
\label{fig:UM323_prop}
\end{figure*}

\subsection{Ionized gas}

We identify emission lines of hydrogen, oxygen and sulphur in the
observed wavelength region. The [O{\sc i}] (8446.4\,\AA) is rather
weak in our spectra, while [S{\sc iii}] at 9068.6\,\AA\ is
prominent. The strength of the Paschen lines can vary and it is
strongest at the peak of luminosity (the most prominent star-formation
regions). After subtracting the stellar continuum we fit the Paschen
lines series from Pa9 to Pa17 with wavelengths from 9229.0658\,\AA\ to
8467.2989\,\AA\ respectively \footnote{For reference, the full list is
  Pa9 (9229.0658\,\AA), Pa10 (9014.9596\,\AA), Pa11
  (8862.8316\,\AA),Pa12 (8750.5203\,\AA), Pa13 (8665.0656\,\AA), Pa14
  (8598.4381\,\AA), Pa15 (8545.4289\,\AA), Pa16 (8502.5287\,\AA), Pa17
  (8467.2989\,\AA), air wavelengths}, as well as [S{\sc iii}] at
9068.6\,\AA.  We masked these lines during the stellar component
fitting, and we fit them a posteriori with independent Gaussians. The
results from the individual Pa lines were averaged (removing the
outliers). Finally, the center of the profile was
shifted in accordance with the main stellar body. Though we wished to
investigate any possible differences between the heliocentric
velocities of the stars and the gas, we found that the literature
values will agree within only 0.5\,\AA, which is 17\kms at 8500\,\AA.
  The thus produced profiles are presented in
Fig.\ref{fig:gas}.

The dispersion of the emission lines is around 30\,\kms and
their rotation is of the same order of magnitude as the \hone\ gas
(see next Section). The Pa and [S{\sc iii}] kinematical profiles
agree well with each other, though the results from the Pa lines are
more noisy. This is due to the lower intensity of these lines and
the limited quality of the data (sky lines, medium S/N),
which can propagate errors in the fitting process, like fitting spikes
instead of the lines.

The wiggles in the velocity field of Mk\,900 (both major and minor
axes) coincide with the similar "wiggles'' in the stellar velocity
profile. This is an indication that the kinematically separate
structures are also present in the gas. For Mk\,324 the Pa lines show
similar profiles as the stars. The \sthree\ line could not be nicely
fitted at the strongest burst, because it was either absent (among the
noise) or very weak, for the rest \sthree\ and Pa emissions follow the
general shape of the stars, rotating in a solid body fashion around
the center of the minor axis. On the major axis of UM038, the ionized
gas displays weak rotation (about 10\,\kms) and one can see the slight
flattening at the center, possibly produced by averaging a solid body
like rotation. The ionized gas along the minor axis rotates with an
amplitude of about 40\,\kms, which is at odds with the offset center
for the stellar rotation. In other words, according to the stellar
kinematics, our slit must have been offset from the rotation axis (the
stars are only approaching us). It means that the rotation axes of the
stars and the gas in UM038 do not coincide. In UM323 the gas does
not rotate along the minor axis while it clearly rotates along the
major axis. We will discuss the similarities and the differences
between the stellar and gas dynamics in the following section.

\section{Discussions}
\label{sect:discussions}
\subsection{Asymmetric drift correction}

The observed velocities of galaxies are slowed down with respect to
the circular velocity by asymmetric drift when significant dispersion
is present. Stars which move on non-circular orbits will spend more
time on their apogalacticon, where they have the slowest rotation.
\cite{binney2008} derived the following correction:
\begin{equation}
v_{c}^{2} = v_{\phi}^{2} +
\sigma_{\phi}^{2}
\left[\frac{\partial\rho}{\partial r} -
  \frac{\partial\ln\sigma^2}{\partial\ln r} 
+  (1 - \frac{\sigma^2_{r}}{\sigma^2_\phi}) 
+ r\sigma^2_{rz}\frac{\partial
    \ln \sigma_{r}^{2} }{\partial \ln r} 
\right].
\end{equation}
Starting from the above formula, we followed \cite{hinz2001} and derived a
simplified formula to correct for the asymmetric drift:

\begin{equation}
\label{eq:asym_drift}
v_{c}^{2} = v_{\phi}^{2} + 2\sigma_{\phi}^{2}\frac{r}{r_{d}},
\end{equation}

where $v_{c}$ is the stellar velocity corrected for asymmetric drift,
$v_{\phi} = v_{obs}/\sin(i)$ is the azimutal velocity (observed
velocity, $v_{obs}$, corrected for inclination, $i$), $r/r_d$ is the
radius normalized to the exponential scale length, and $\sigma_{\phi}^2
= \frac{\sigma_{obs}^2}{2cos(i)^2+sin(i)^2}$ is the observed velocity
dispersion ($\sigma_{obs}$) corrected for inclination.  To derive the
scale length of our galaxies we divide the effective radius by 1.678
\citep[][for exponential profiles]{prugniel1997}.  The approximations
we use to derive Eq.\,\ref{eq:asym_drift} are the following: (1) the
spatial density of stars is assumed to be exponential (as in spiral
galaxies); (2) we neglect $\langle v_rv_z \rangle$, which is correct
only for relaxed objects, while our objects may not be relaxed yet;
(3) looking at the \hone\ velocity fields (Fig.\ref{fig:gas_stars}) we
assume solid body rotation, thus $\sigma^2_{r}/\sigma^2_\phi = 1$; (4)
finally, no change of the radial velocity dispersion along the radius
is imposed by neglecting the $\frac{\partial \ln \sigma_{r}^{2} }{\partial
  \ln r}$ term.

The velocity profiles corrected for inclination and asymmetric drift 
are shown on Fig.\,\ref{fig:gas_stars} with blue circles. Though
these corrected velocities are more in line with the \hone\ velocities,
it is clear that the stars are over-corrected since they have higher
velocities at a given radius than the \hone\ gas. This is not surprising
since the correction was oversimplified to make it computable.

\subsection{Stellar vs. Gas kinematics}
\label{subsec:star_gas}

One could expect that stars and gas in a galaxy have similar
kinematical characteristics, since the former are a product of the
latter. However, nature proves to be more complex than this \citep[see
for example][]{ostlin2004}. Strong stellar and supernova winds can disturb
the gas around a star-formation region, or the stars could be
preferentially formed from low angular momentum gas, that fell towards the
center of the galaxy \citep{elemgreen2012}. External factors
also disturb the galactic dynamics, such factors may include mergers
or ram-pressure stripping. The latter tends to sweep the gas off the
galaxy but does not have much effect on its stars, while the former
can be a rather violent event for the host galaxy, combined with
star formation and producing irregularities in the gas and stellar
kinematics \citep{verbeke2014,cloet-osselaer2014}.

Nevertheless, when comparing different classes of dwarf galaxies and
investigating the possible connection between them, the gaseous
kinematics of star-forming dwarfs (dIrrs, BCDs) are often compared to
the stellar kinematics of quiescent dwarfs, like dwarf ellipticals
\citep[dEs, e.g.][]{vanzee2001}. The reason for this is obvious - dEs
contain little if any gas, while accessing the stellar population
characteristics of dIrr or BCD galaxies is challenging because of the
overpowering gas and young stars emission.

\begin{figure*}
\includegraphics[width=0.40\textwidth]{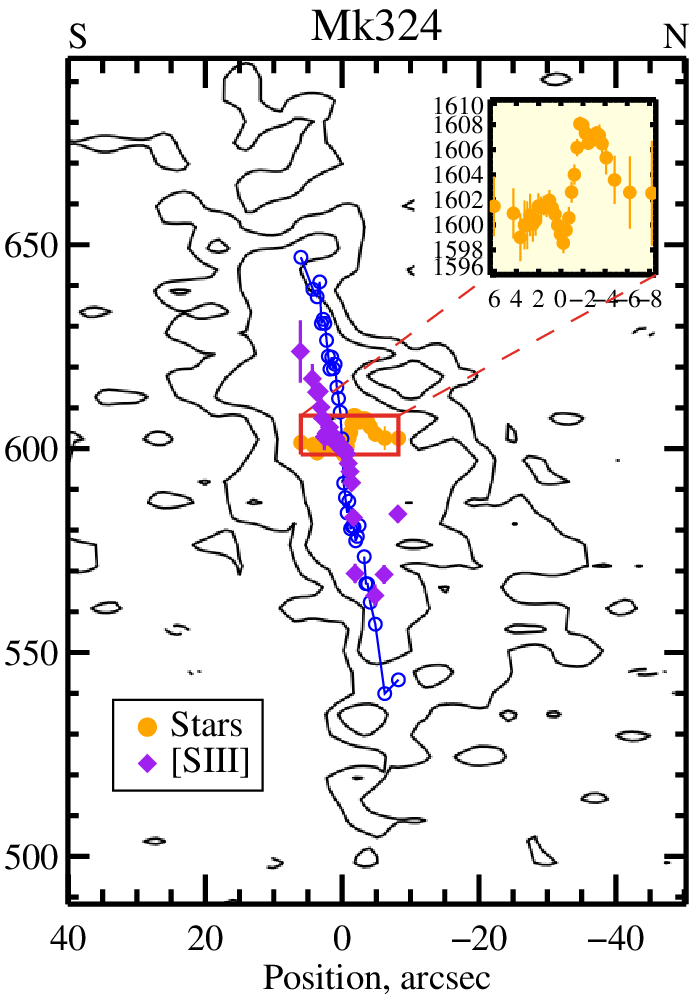}
\includegraphics[width=0.53\textwidth]{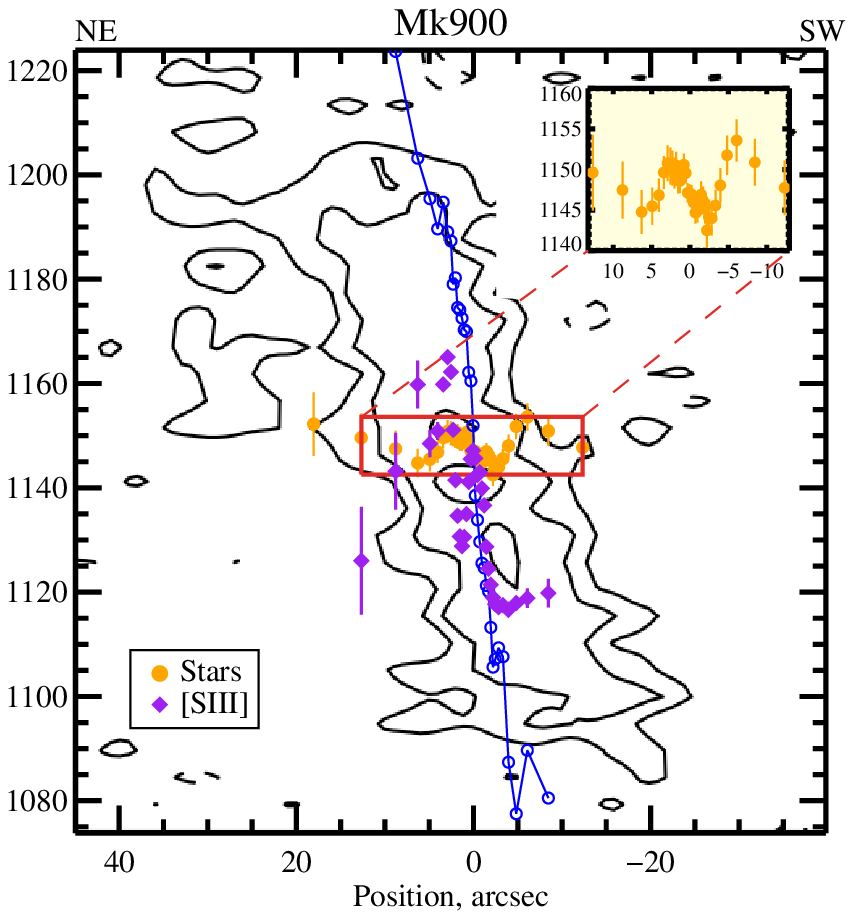}
\includegraphics[width=0.53\textwidth]{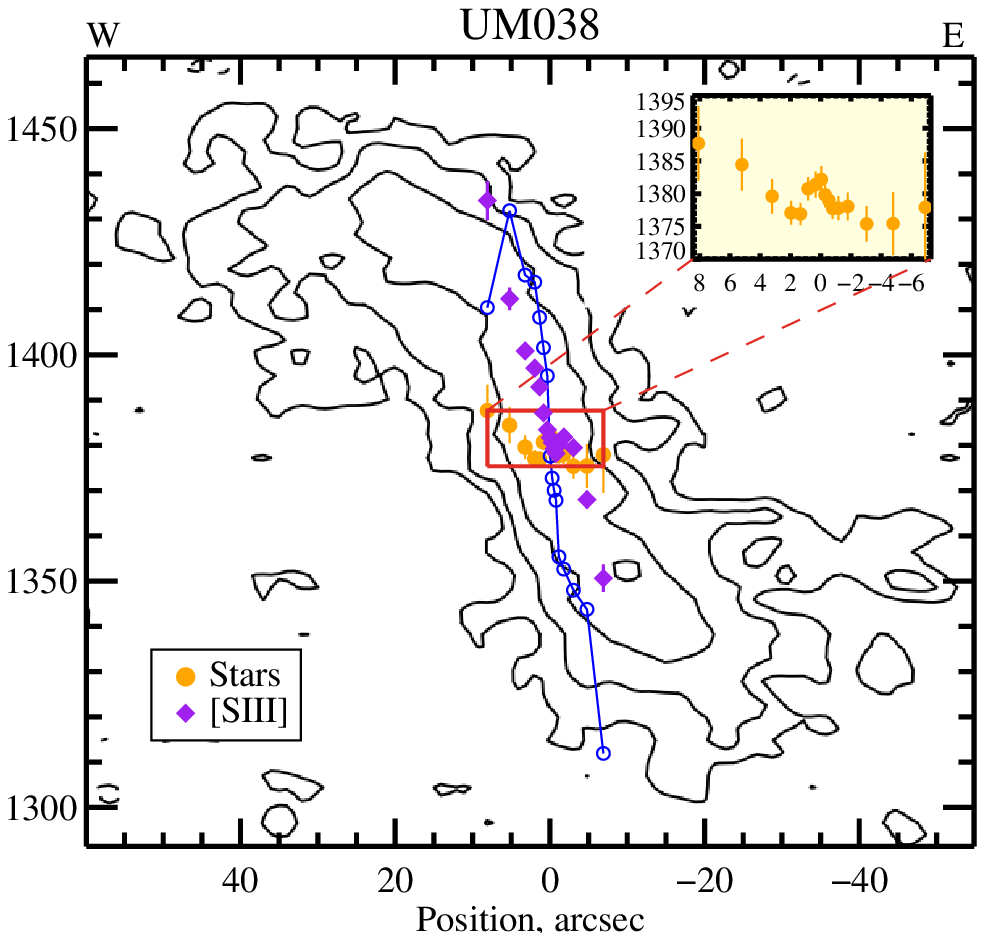}
\includegraphics[width=0.40\textwidth]{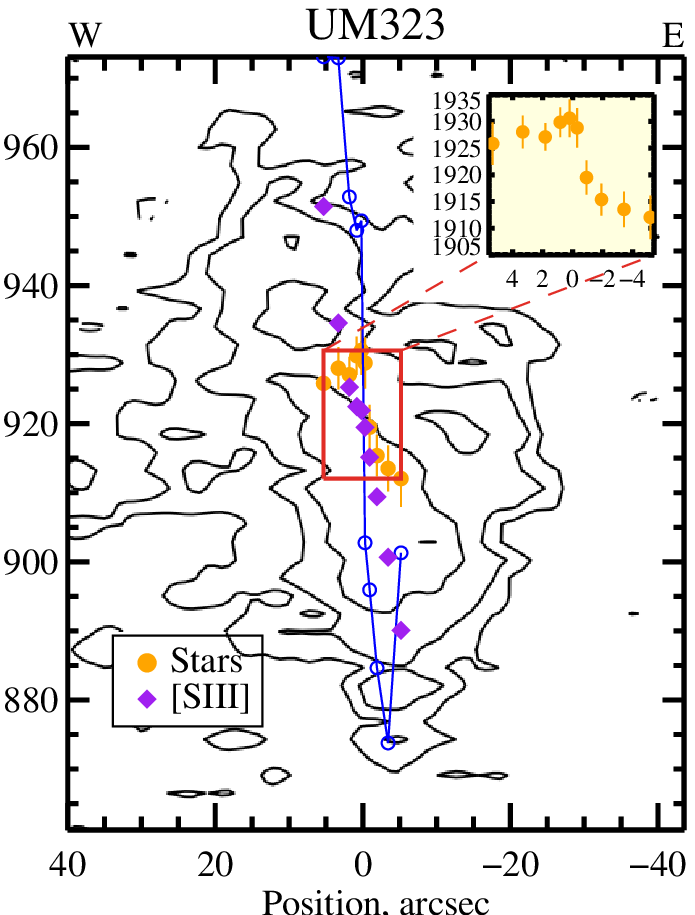}

\caption {Radial distribution of the neutral gas \citep[contours,
  representing $-3\sigma, 3\sigma, 6\sigma$, and $12\sigma$,
  from][their fig. 12]{vanzee2001}, ionized gas (purple diamonds) and
  stellar (orange circles) velocities of our sample. The open blue
  circles indicate the stellar velocities corrected for asymmetric
  drift. Our measurements are smoothed with a boxcar running average
  of width 2. The zoomed regions at the upper right corners show the
  central stellar velocities. }
\label{fig:gas_stars}
\end{figure*}

Here, however, we derived the stellar and ionized gas kinematics of
four BCDs and we compare them (Fig.\,\ref{fig:gas_stars}) with \hone\
kinematics from the literature \citep{vanzee2001}. We selected the
(mirrored) major axes which are the same as the one chosen from the
\citeauthor{vanzee2001} positional angles (PA) for maximum \hone\
rotation (their fig.\,12).
We smoothed our values with a boxcar running average with a width of
two. Thus, we could suppress some local ``jumps'' resulting from the
observational errors, and we can follow the overall behavior.

Despite the relatively different photometric appearances of our
galaxies they all seem to follow the same kinematic patterns. The
stellar body seems not to rotate or rotates only with a few \kms. All
of the stellar velocities experience central disturbances, indicating
a presence of a disk-like structure. The ionized gas mostly follows
the \hone, except for the center, where it is following the central
stellar disks. In the case of Mk\,900 the ionized gas velocity field,
seems to have the same pattern as the stars, but with higher
amplitudes. The stellar velocities corrected for asymmetric drift ($v_c$) are
more consistent with the \hone\ rotation, but once again they are
clearly over-corrected as they will indicate a stellar body rotating
faster than the neutral gas.

These observations lead us to speculate that the stellar burst must be
triggered by a process which can change/disturb the stellar and gas
kinematics, such as an in-falling gas cloud or in-spiraling star
formation clumps (see Sect.\,\ref{subsec:triggers}).  However, some of
the star-forming regions have a significantly higher metallicity than
the galaxy body, a peculiar rotation velocity, and a higher age
indicating that they can remain stable for several Gyr, thus hinting
at the presence of dark matter associated with these regions.

\subsection{Burst triggering mechanisms}
\label{subsec:triggers}

The unusually high star-formation rates in BCDs have been contributed
to several processes, including: in-spiraling gas clumps
\citep{elmegreen2012}, cloud impact(s) \citep{gordon1981}, mergers
\citep{bekki2008}, tidal effects \citep{vanzee1998}. 
Though Mk324 and UM323 have relatively close companions and 
interactions cannot be fully excluded it is nevertheless unlikely given that
these objects are relatively faint and small (see
Sect.\,\ref{subsec:phot} and Fig.\,\ref{fig:imaB}). 

Relatively massive star-formation regions in the galaxies can lose or
change their angular momentum by torques with the other galactic
components \citep{elmegreen2012} and spiral into the center of the
galaxy in analogy to the high-$z$ formation of bulges
\citep{elmegreen2009}. This process could explain several of the
(unusual) characteristics of BCDs like the high star-formation rate,
the dense stellar and gaseous cores (compared to dwarf irregulars)
and the differential rotation observed in some BCDs. However, in our
sample we have mostly solid body rotation for the \hone\ and ionized
gas, and the stars corrected for asymmetric drift (see
Fig.\,\ref{fig:gas_stars} and \citealt{vanzee2001}) wich makes
friction unlikely \citep{goerdt2010}, except in the outer parts of
UM038. On the other side, the bursts in most of the galaxies seem to be
relatively longly lived ($\gtrsim 1$\,Gyr), which can be explained by
rejuvenating in-spiraling clumps \citep{elmegreen2012}. This process
could also explain the star formation regions with both higher and
lower metallicities in comparison to the rest of the galaxy, since the
metal abundance will depend on the original birthplace of the
clumps (assuming chemical inhomogeneity of these galaxies).

Another plausible mechanism, which could explain the star burst in the
4 galaxies in our sample, is an impact with a gas cloud. We run
self-consistent, N-body/SPH simulations of impacts between a dwarf
galaxy and a gas cloud \citep{verbeke2014}. We chose the gas clouds
with masses and densities in accordance with observations of high
velocity clouds \citep[][for a review]{wakker1997}. We find that such
events can indeed trigger a burst of star formation, enhancing the
normal star formation level by a factor of ten.  Our simulations can also
produce star formation regions with different metallicities, depending
on whether the star formation material is simply compressed by the incoming
gas cloud or if it is the new gas which forms the stars. They could
also explain bursts living for about a Gyr, because of the
self-induced nature of the star formation - the winds of new stars and
supernov\ae\ expel but also compress the gas, provoking new star
formation in these condense regions \citep{gerola1980}. However, having
regions with star
formation lasting more than a Gyr - which are self-enriched, and
have peculiar kinematics - seem to require a presence of dark
matter substructures \citep{cloet-osselaer2014}.

Another
difficulty assuming such a scenario is coming from the relatively
special conditions for producing a burst. The burst ignition depends a
lot on the properties of the gas cloud and the host galaxy, like the
density of the gas cloud, the trajectory, and the mass of the host
dwarf galaxy.  On the other side, BCDs are also relatively rare, about 100 times
less than other dwarfs with $B$-band magnitude around $-16$\,mag. 
\citep[][their fig.\,6]{sanchez-almeida2008}.

\subsection{Dynamical masses}

Making the assumption that our galaxies are dark matter dominated and
that they lie in a spherically symmetric dark matter potential, we can
estimate their dynamical masses using 
the drift corrected velocities from the previous section: 
\begin{equation}
M(<r_{eff}) =v_c^2r_{eff}/G,
\label{eq:mass_vel} 
\end{equation}
where $G$ is the gravitational constant ($G =
4.3\times10^{-3}$\,pc\,M$_\odot^{-1}$), $v_c$ is the inclination and
asymmetric drift corrected maximum velocity at radius $r$.

Because our galaxies are not thin disks for which the asymmetric
drift correction approximations are not fully valid and because we do
observe very little stellar rotation, we also estimate
the masses from the velocity dispersion of the stars, which is
significant. We use 
the following formula:
\begin{equation}
M(<r_{eff}) = \mu r_{eff} \sigma^2,
\label{eq:mass_sig}
\end{equation}
as in \cite{walker2009}, where $\mu$ is 580\,M$_\odot$/pc/(\kms)$^2$,
$\sigma$ is the mean stellar velocity dispersion (see Sect.\,
\ref{sect:results}) and $r_{eff}$ is the effective radius. Both
sets of estimated masses are reported in
Table\,\ref{table:masses}. The two masses are consistent within a
factor of 1.3. 
It is worth noticing that the discrepancy between the two estimations
comes from the fact that these galaxies are probably not relaxed,
which is also visible from the disturbed \hone\ \citep{vanzee2001}, and
we cannot use the Virial theorem.

We also compute the dynamical $M/L$ ratio using the $\sigma$-derived
dynamical masses and the $I$-band luminosities
(Table\,\ref{table:phot}).  Our galaxies lie between
0.8\,(M$_\odot$/L$_\odot$) and 2.4\,(M$_\odot$/L$_\odot$).

\begin{table}
  \caption{Dynamical masses of our galaxies. In the first column we print the name of the
    galaxy, in the second column the effective radius in $I$-band, third column is the mass
    derived from the asymmetric drift corrected velocity  according to
    Eq.\,\ref{eq:mass_vel}, in the forth column we print the
    dynamical mass derived with the luminosity weighted velocity
    dispersion according to Eq.\,\ref{eq:mass_sig}, the last column
    (5) is the dynamical mass-to-light ratio.}
\begin{tabular}{lrrrr}
\hline
Name & r & M$_v$ &  M$_\sigma$  & M$_{\sigma}$/L$_I$ \\
& (kpc) & (10$^{8}$ M$_\odot$) &  (10$^{8}$
M$_\odot$) & (M$_\odot$/L$_\odot$)\\
\hline
\input{dyn_mass.tex}
\hline
\end{tabular}
\label{table:masses}
\end{table}

Using the stellar mass-to-light ratio (derived from the stellar
population parameters) and the dynamical mass-to-light ratio we can
estimate the $M_{dyn}/M_\star$. We find it to be more than two and
conclude that if our objects are relaxed they are dark matter
dominated in the inner R$_{eff}$.  \cite{vanzee2001} also estimated
the dynamical masses of these objects from the \hone\ velocity fields,
which reach several times further than our stellar velocity
profiles. The masses reported by these authors are about an order of
magnitude higher than ours because they trace the solid body rotation
curves further outwards than us.

\subsection{Comparison with early type dwarfs}

Our sample was chosen for its resemblance to dwarf elliptical galaxies
as a possible end product of the BCDs after the gas is
exhausted/removed and the burst is quenched. In this section we will
investigate if such a transition is possible.

We derived the spatial SSP-equivalent ages and metallicities of our
sample of BCDs. While these values are typical for star forming dwarfs
with low metallicities and relatively young ages, we anticipate that
these are not young objects and that the bulk of their stellar mass
was formed a long time ago (extended stellar halo is visible in our
$I$ and $J$ band images, Fig.\,\ref{fig:imaI},\ref{fig:imaJ}). Though
we could not perform star formation history analysis, due to the
limited wavelength range, we anticipate that if their gas is removed
and star formation ceased, only after 1-2 Gyr of passive evolution
their stellar populations will be identical to early type dwarfs, as
shown from other observational studies
\citep{weisz2011,sanchez-almeida2008, koleva2013}.

\begin{figure*}
\includegraphics[width=0.98\textwidth]{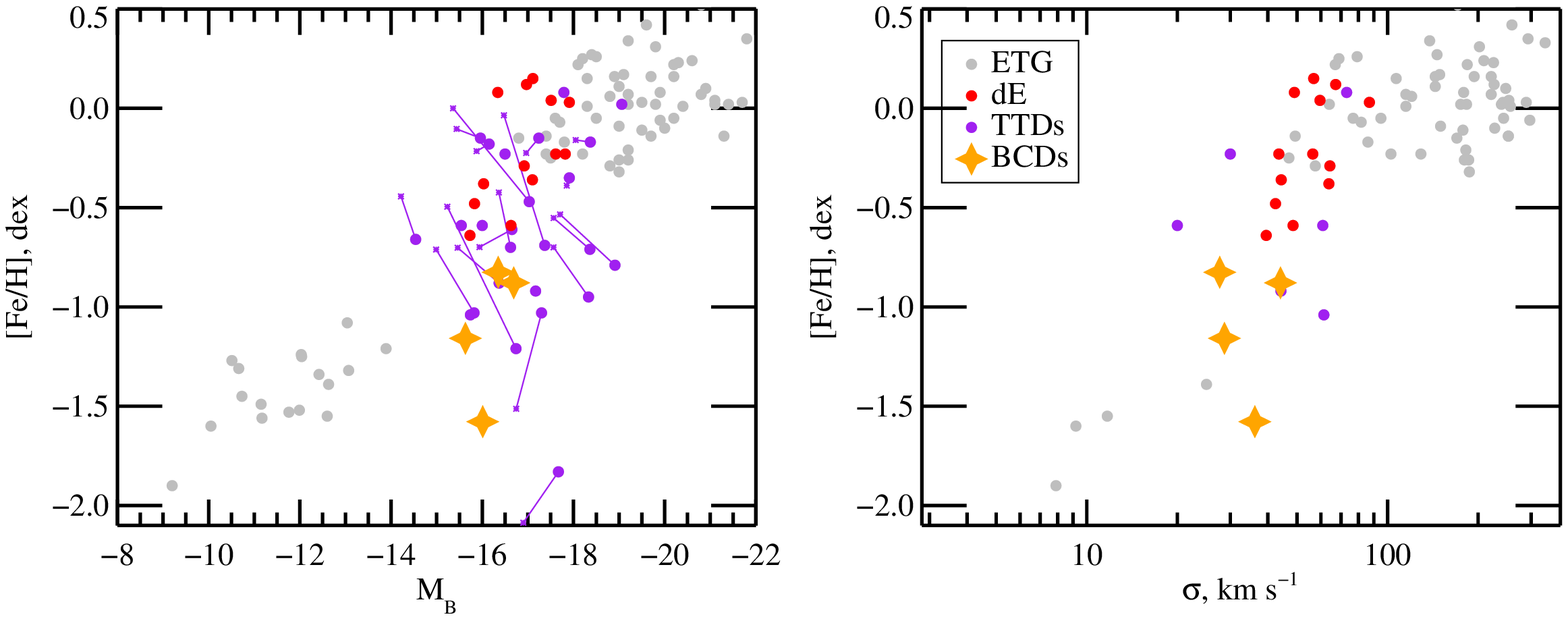}
\caption { Metallicity - luminosity and metallicity-velocity dispersion
  relations for a set of early type galaxies (ETG, gray circles) from
  \citet[][and the references therein]{koleva2011}. With red circles
  we denote dwarf elliptical galaxies from \citet{koleva2009}, while
  purple dots are transition type dwarfs \citep{koleva2013}. We also
  show the locus of the studied transition type dwarfs after 1\,Gyr of
  passive evolution (small purple stars).
}
\label{fig:mass_met}
\end{figure*}

The final metallicity of a stellar system, or how much metals are
going to be produced and kept in a galaxy, depends on its total mass,
density, geometric configuration, and star formation history. Thus,
the mass-metallicity relation can in principle indicate the
evolutionary stage of a stellar system and its star formation modes
\citep[e.g.][]{tremonti2004}.  To check if our sample of BCDs had
similar star formation histories as other early type galaxies (ETGs)
we place them on the mass-metallicity
relations. Fig.\,\ref{fig:mass_met} shows the relation between the
metallicities of these objects as a function of their velocity
dispersion and $B$-band magnitudes. The data for dwarf ellipticals and
transition type dwarfs are taken from \cite{koleva2009,koleva2013},
where the metallicities are the SSP-equivalent central metallicities
of the objects. Since the star formation regions of our BCDs do not
coincide with the geometrical center of the galaxy we took the mean
metallicities, they are reported in Table\,\ref{table:spec} and indicated on
their radial profile plots
(Fig.\,\ref{fig:Mrk324_prop}--Fig.\,\ref{fig:UM323_prop}). Our sample
seems to occupy the same locus as the transition type galaxies in the
metallicity-velocity dispersion plane, while the BCDs $B$-band
luminosities are slightly too bright for their metallicities than the
quiescent dwarfs. Since these BCDs are forming stars at extraordinary
rates (around $0.1$\,\msol/yr \citealt{vanzee2001}), their $B$-band
magnitudes are understandably higher for their metallicities. Thus, in
this case, the velocity dispersion is probably a better indicator for
the enclosed total mass of the system (though our galaxies may not be
dynamically relaxed).

Another ground of comparison is the dynamics of dwarf elliptical and
blue compact dwarf galaxies. While their difference in the v/$\sigma$
values were long considered as an obstacle
\citep[e.g.][]{vanzee2001}, we have shown here that when using
the stellar rotations and velocity dispersions values our galaxies
have (v/$\sigma$)$_\star \sim 1.5$, which is perfectly in the range of
dwarf elliptical galaxies, that lie between (v/$\sigma$)$_\star $
0.1 and 2.5 \citep{toloba2011}. Moreover, it has been found that the
surface brightness profile of Virgo cluster BCDs is similar to the
surface brightness profile of some of the dEs there
\citep{meyer2013}, which may suggest similar gravitational
potentials. Moreover, \cite{lelli2013} found that the velocity gradient
(velocity as a function of the radius) in
BCDs and dEs is similar and reflects the distribution of the central
total mass. 

Our BCDs, though being selected to look as regular as possible and
resemble dEs, have very disturbed velocity fields. Mk324 is the best
example where the body is relaxed and regular while the blue clump to
the NW of the galaxy has very different kinematics - this structure
seems to be kinematically decoupled from the rest of the galaxy, with
a much smaller velocity dispersion and even a velocity gradient of a few
\kms. The embedded structure we observe in the stellar rotational
curves of our targets are also present in high resolution observations
of dEs especially if 3D observations are available
\citep{geha2005,chilingarian2007, rys2013}. We expect, however, most
of these substructures to disappear with the ceasing of the star
formation and the relaxation of the systems.

\section{Conclusions}
\label{sect:conclusions}

We have selected four blue compact dwarf galaxies with relaxed and
regular outer regions, resembling dwarf elliptical galaxies. They are
part of a sample of 6 BCDs observed from \cite{vanzee2001} in the
radio domain. We have obtained $B$, $I$ and $J$ images and long-slit
data along their kinematical major and minor axes
\citep{vanzee2001}. The spectroscopic data were obtained near the CaT
region to ensure more relaxed kinematical profiles and milder age
sensitivity. From the photometric data we have obtained total magnitudes
and half light radii. They span a range of $-16.74 > M_I > 18.17$\,mag
and their $I$-band effective radii are of the order of 0.5\,kpc.
Their ($B-I$) colours in the star formation regions are close to
$-0.5$, typical for young population. The bodies of the galaxies have
($B-I$) colours around 1.5. Most of the strongest star formation
regions are also detected in the $J$-band images, indicating
that the burst is not a recent event.

We analysed the spectroscopic data using \ulyss\ - a full spectrum
fitting package together with Pegase.HR/CFLIB population synthesis
models. The CaT region is not very sensitive to the age of a stellar
system, but can still deliver reliable ages and metallicities,
providing that these quantities are below a few Gyr and $-0.5$\,dex,
respectively.  We found population ages around 1\,Gyr to dominate the
light. The metallicities of these stars are 1/10$^{th}$ of the solar
value.

The velocity profiles of our galaxies are complex, with disk-like
rotation in the regions
of active star formation. These are also 
often associated with drops in the velocity dispersion. 

Considering all the observed properties we conclude that there is no
unique mechanism which triggers the bursts, even in the same galaxy. In
some cases (Mk324 and Mk900) the cold dynamical structure, the
relatively high metallicity and the long life span of star formation
region point towards mergers \citep{bekki2008,cloet-osselaer2014}. In
other cases (UM038), when differential rotation is detected, the burst
can be explained with in-spiraling gas clumps \citep{elmegreen2012}.
Finally, cloud impacts \citep{gordon1981,verbeke2014} are possible
ignitors when the metallicity of the star-formation region is low, the
velocity field is disturbed and the life span of the burst is
relatively short.

When including the effect of asymmetric drift our stellar velocity
curves coincide with the velocities curves of the ionized gas (Pa and
\sthree\ lines) and the \hone\ data from \cite{vanzee2001}. The
velocity profiles show a rapid increase in the inner 1\,kpc of the
galaxies in a solid body fashion, indicating dense cores. Otherwise,
our values of the stellar ($v/\sigma$)$_\star$ are spread between
0.4 to 2.8, so they are in the range of those for dwarf
elliptical galaxies \citep{toloba2011}.
When placed on the metallicity-velocity dispersion relation our BCDs
occupy similar regions for a given $\sigma$ as the transition type
dwarfs, which in turn are on their way of becoming dwarf ellipticals
\citep{koleva2013}. 

Finally, we computed the dynamical masses of our galaxies inside the
effective radii and we obtain values of the order of a few $\times
10^8$\msol.

\section*{Acknowledgments}

MK is a postdoctoral fellow of the Fund for Scientific Research-
Flanders, Belgium (FWO11/PDO/147) and Marie Curie (Grant PIEF-
GA-2010- 271780).  
We acknowledge the use of 
 SDSS DR7 (the full acknowledgment can be found
on \url{http://www.sdss.org/collaboration/credits.html}) and 
the HyperLeda database (\citealt{hyperleda}, \url{http://leda.univ-lyon1.fr}).
The publication is supported by the Austrian Science Fund (FWF). RB
acknowledges financial support from the 
CHARM framework (Contemporary physical challenges in Heliospheric
and AstRophysical Models), a phase VII Interuniversity Attraction Pole
(IAP) programme organised by BELSPO, the BELgian federal Science
Policy Office.

\bibliographystyle{mn2e}
\bibliography{bcd}   

\appendix

\section{Photometry}
\label{app:photometry}
In this appendix we present the photometric images of our 4 targets in
$B$-band, $I$-band and $J$-band. We plot the full field-of-view of the
instruments (for the $B$ and $I$ images) to appreciate the relative
isolation of the objects. The $J$-band images are mosaics, produced
using a fast pixel-shift algorithm and weight-maps.

\begin{figure*}
\includegraphics[width=0.49\textwidth]{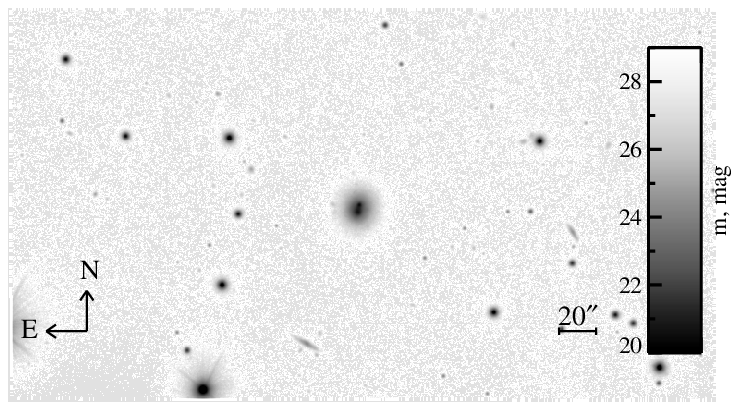}
\includegraphics[width=0.49\textwidth]{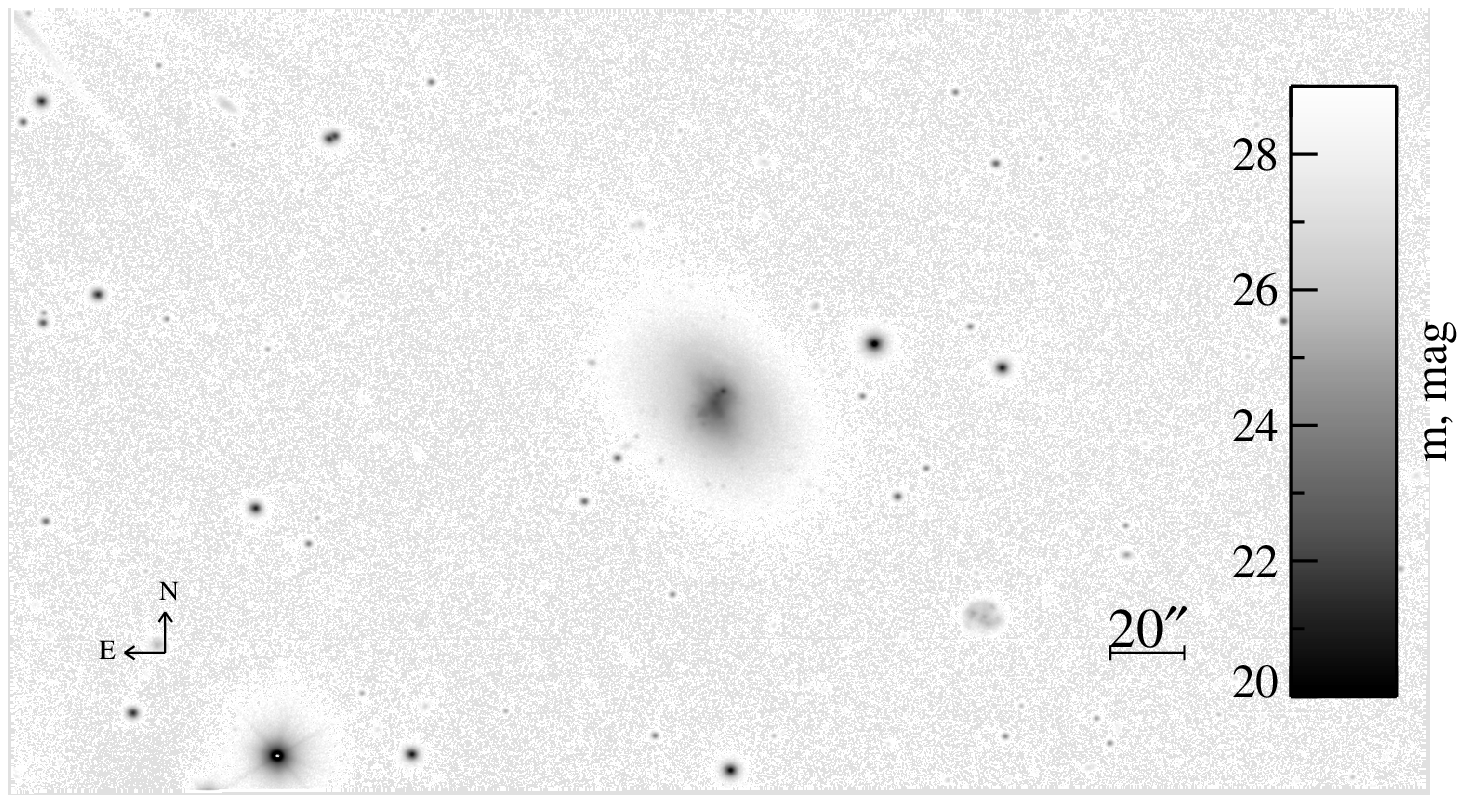}
\includegraphics[width=0.49\textwidth]{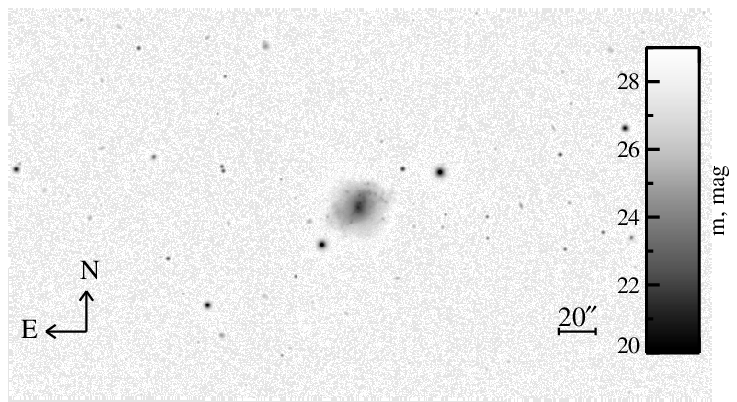}
\includegraphics[width=0.49\textwidth]{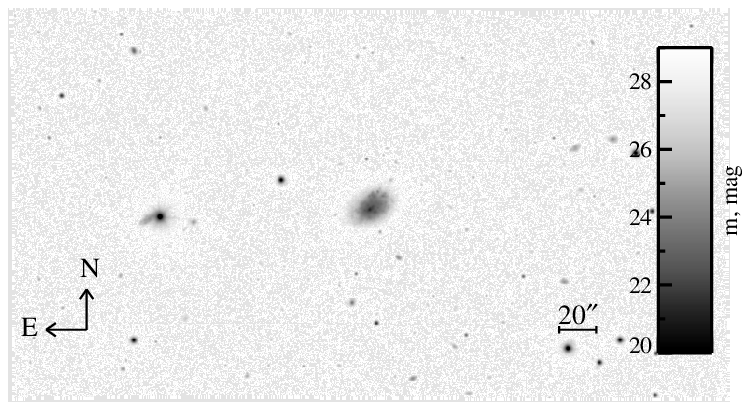}
\caption {From left to right $B$-band images of Mk324, Mk900, UM038,
  UM323. The sizes of the fields are $1.33 \times 1.33$\,arcmin. We
  indicate the scale of each of the images (20\,arcsec line) and the
  NE orientation.}
\label{fig:imaB}
\end{figure*}
\begin{figure*}
\includegraphics[width=0.49\textwidth]{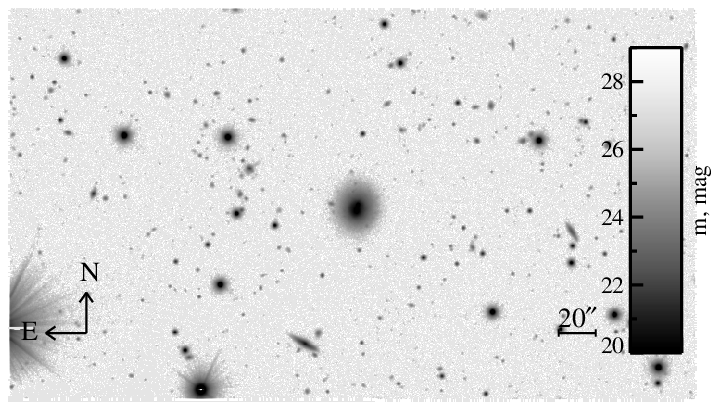}
\includegraphics[width=0.49\textwidth]{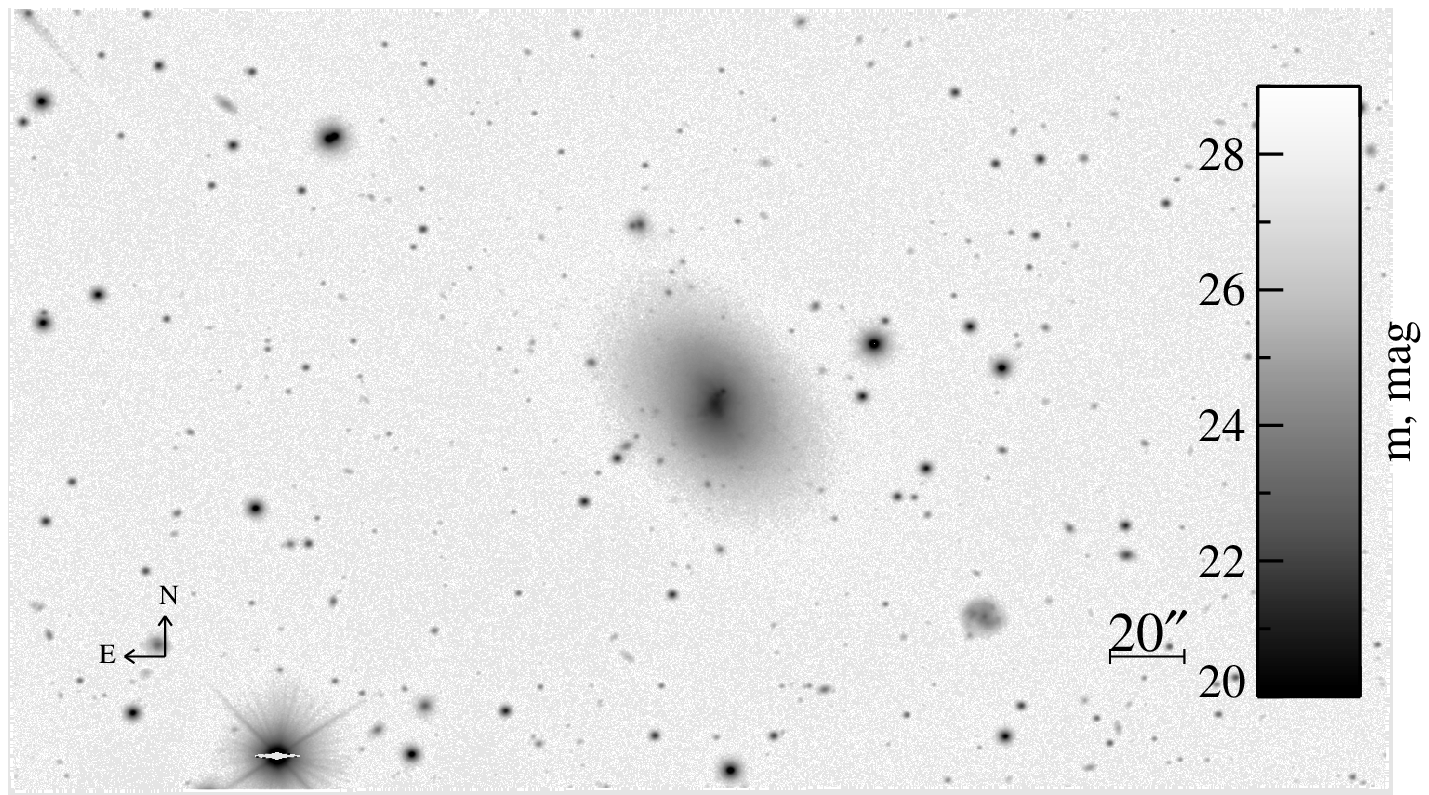}
\includegraphics[width=0.49\textwidth]{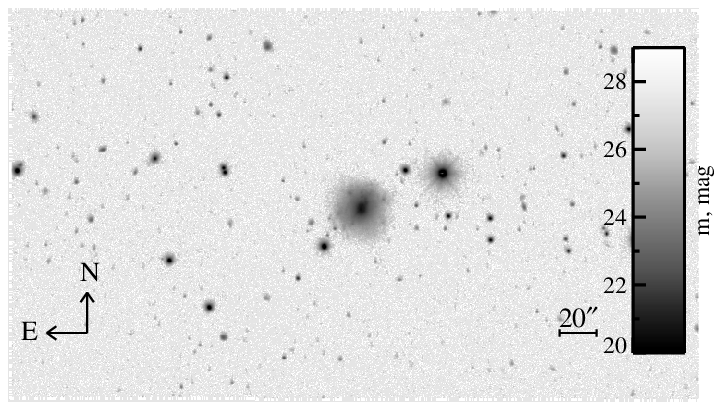}
\includegraphics[width=0.49\textwidth]{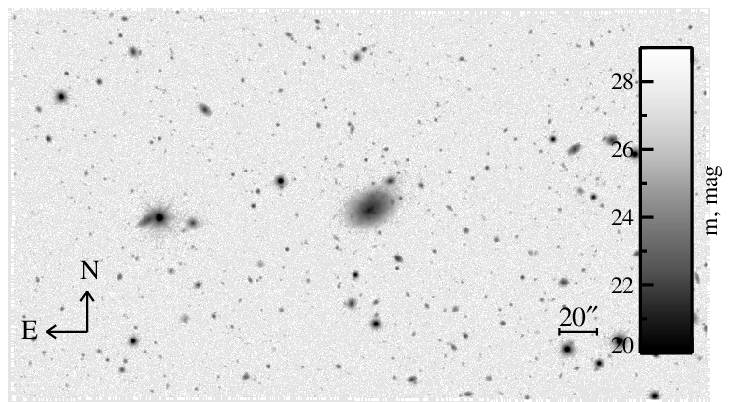}
\caption {From left to right $I$-band images of Mk324, Mk900, UM038,
  UM323. Same indications as in \ref{fig:imaB}}
\label{fig:imaI}
\end{figure*}
\begin{figure*}
\includegraphics[width=0.49\textwidth]{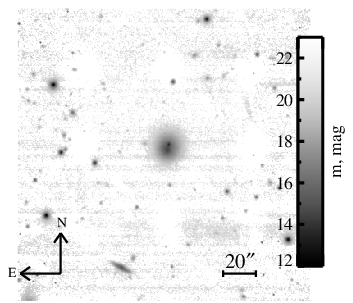}
\includegraphics[width=0.49\textwidth]{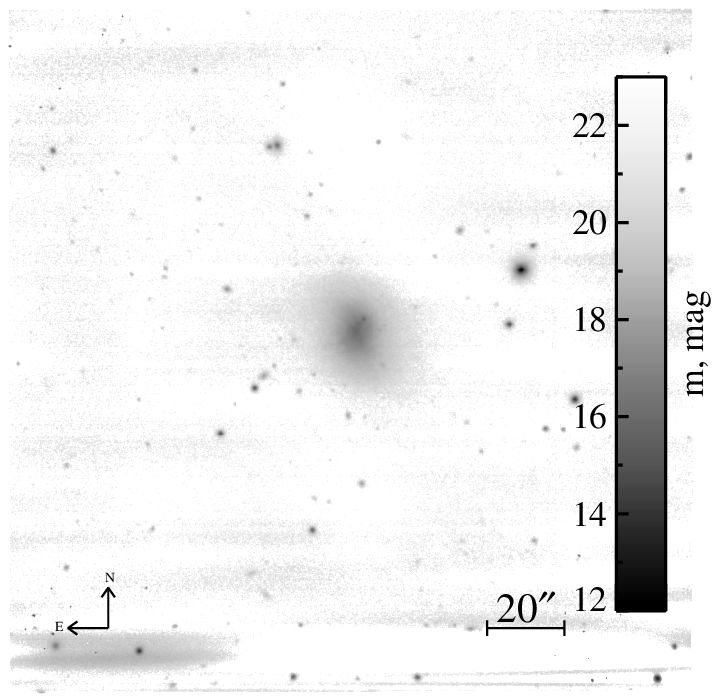}
\includegraphics[width=0.49\textwidth]{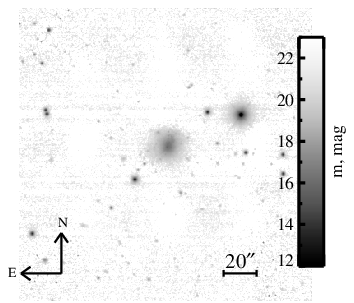}
\includegraphics[width=0.49\textwidth]{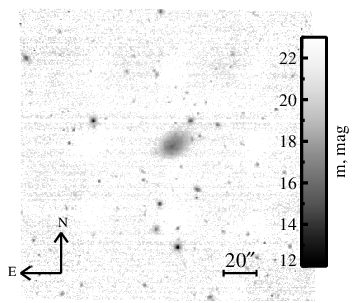}
\caption {From left to right $J$-band images of Mk324, Mk900, UM038,
  UM323. Same annotations as in Fig.\,\ref{fig:imaB}.}
\label{fig:imaJ}
\end{figure*}

\section{Additional figures}

\begin{figure*}
\includegraphics[width=0.24\textwidth]{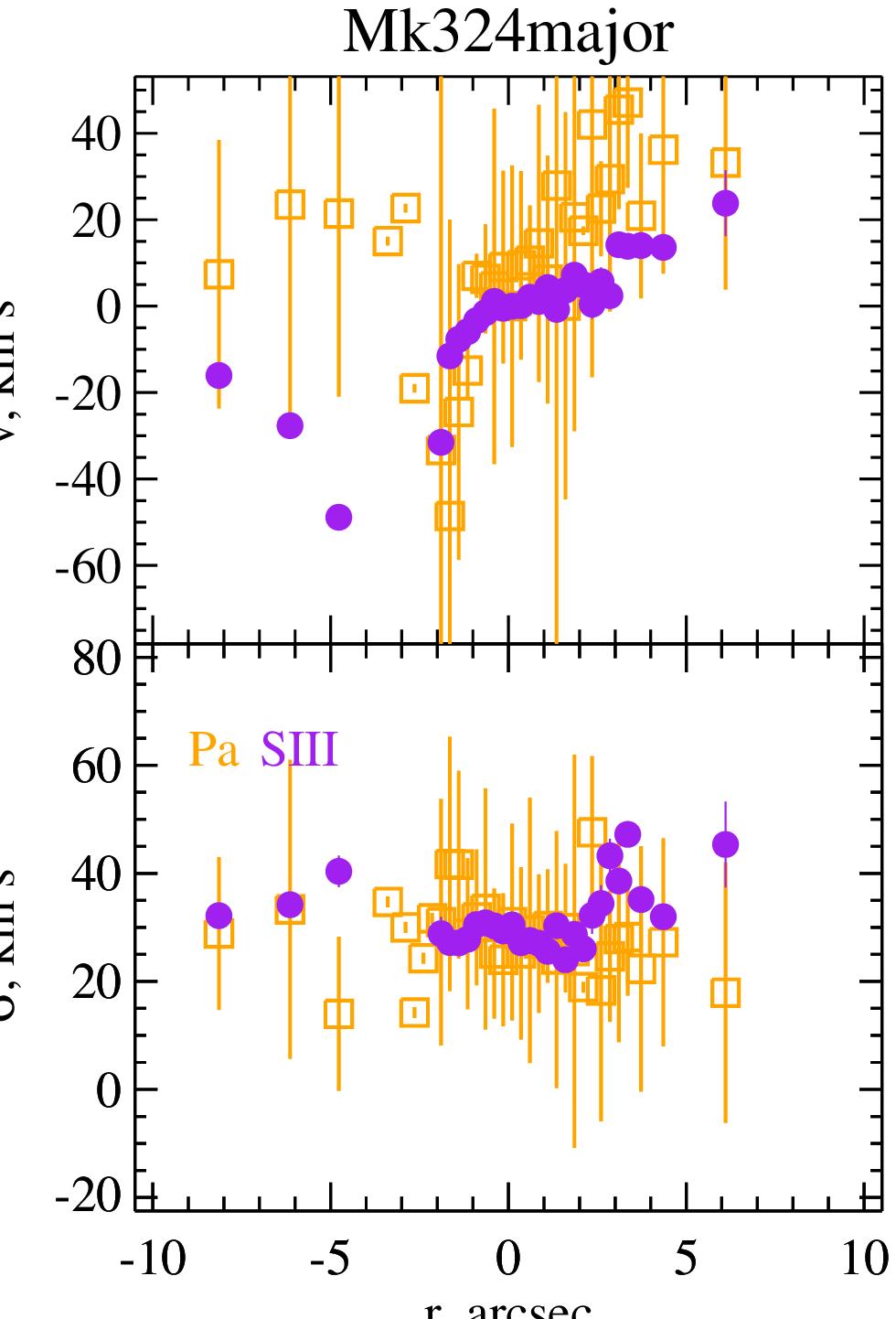}
\includegraphics[width=0.24\textwidth]{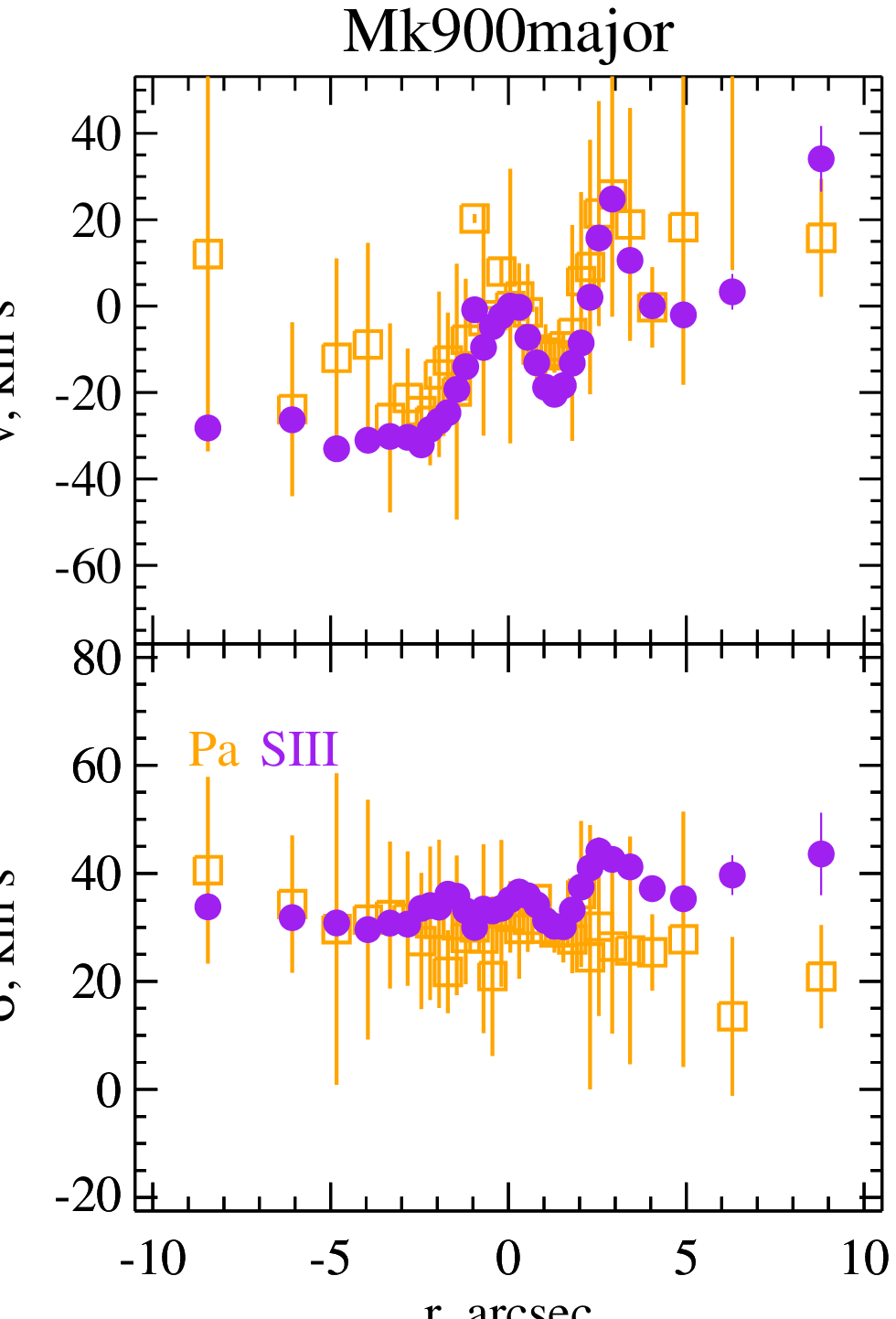}
\includegraphics[width=0.24\textwidth]{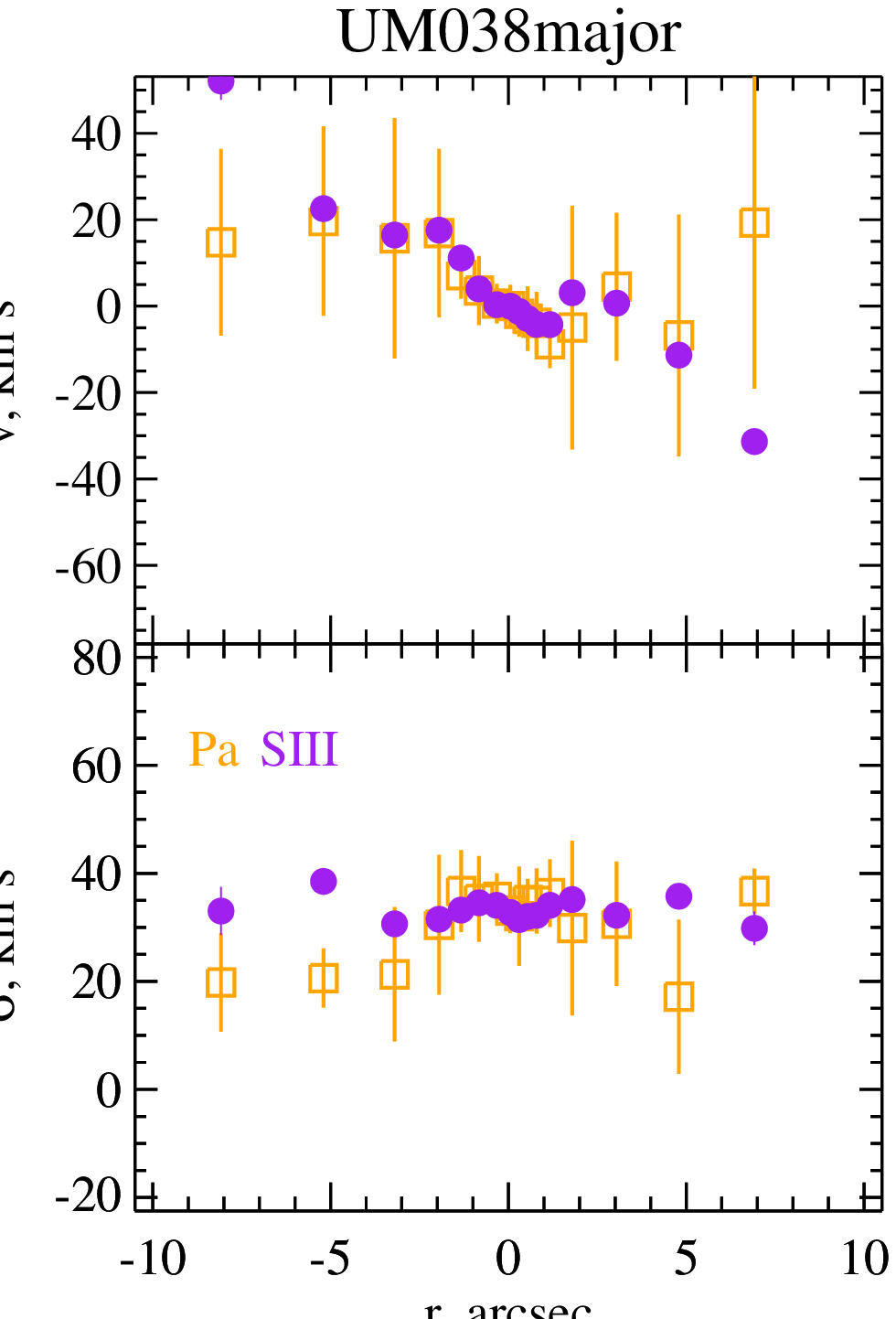}
\includegraphics[width=0.24\textwidth]{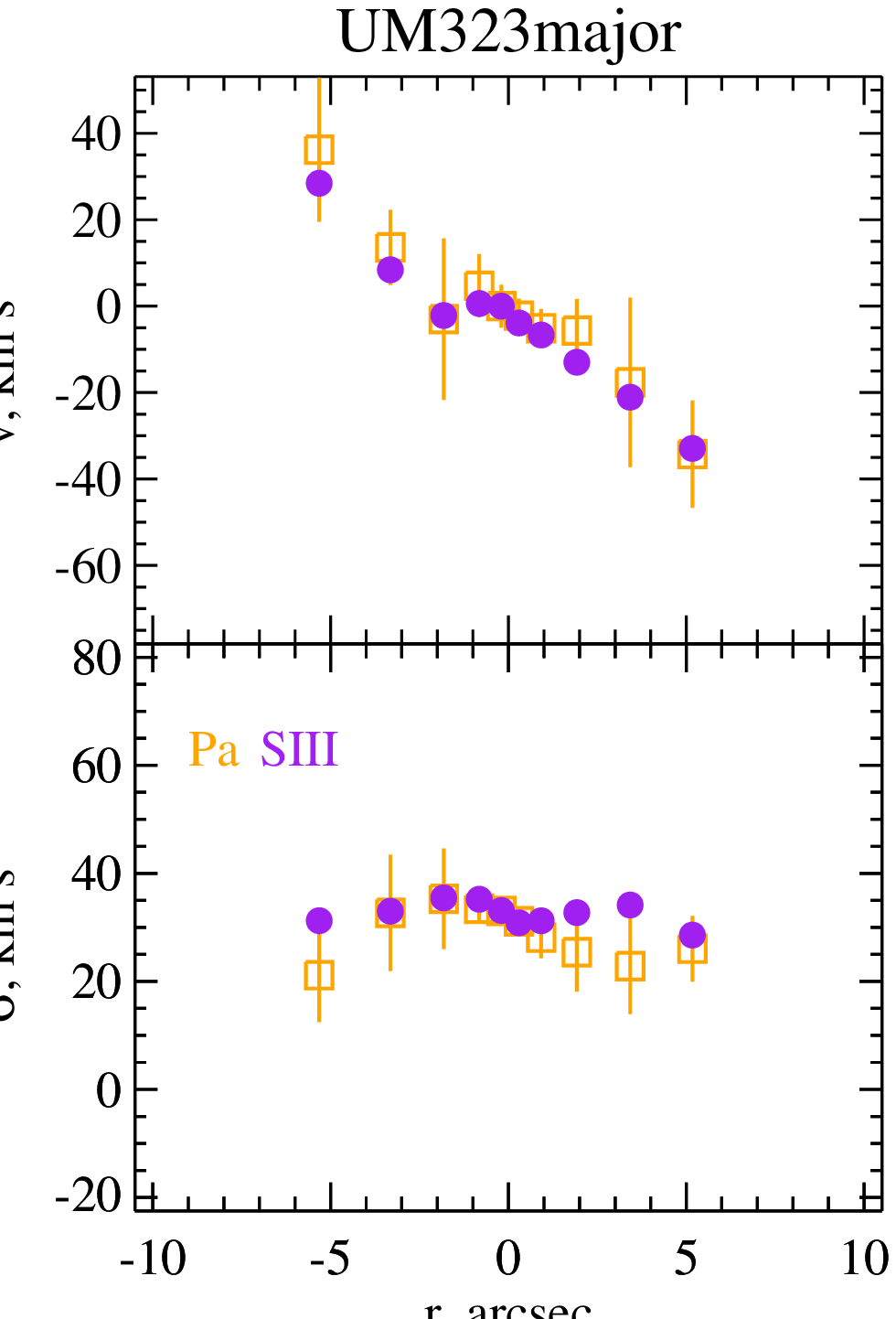}
\includegraphics[width=0.24\textwidth]{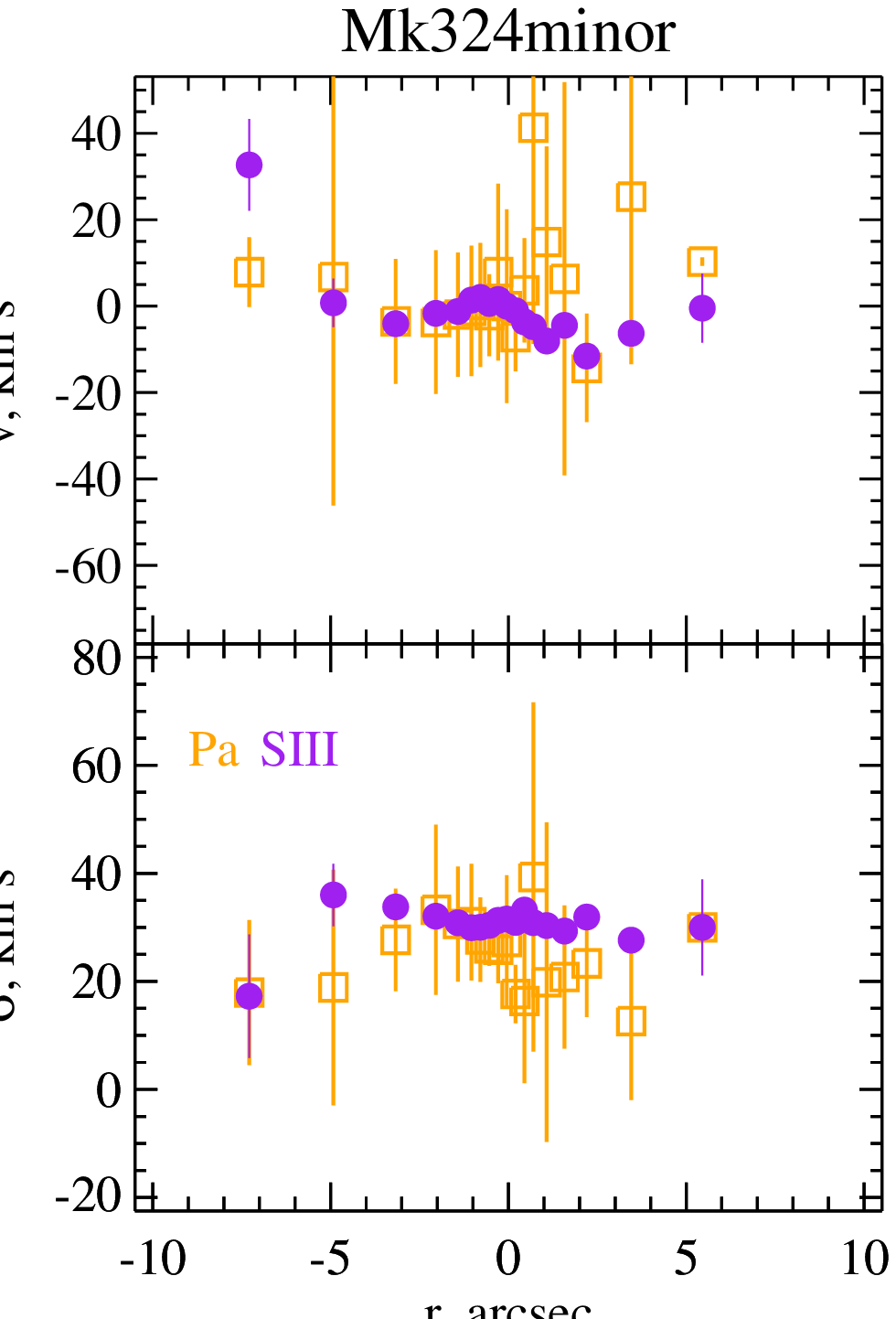}
\includegraphics[width=0.24\textwidth]{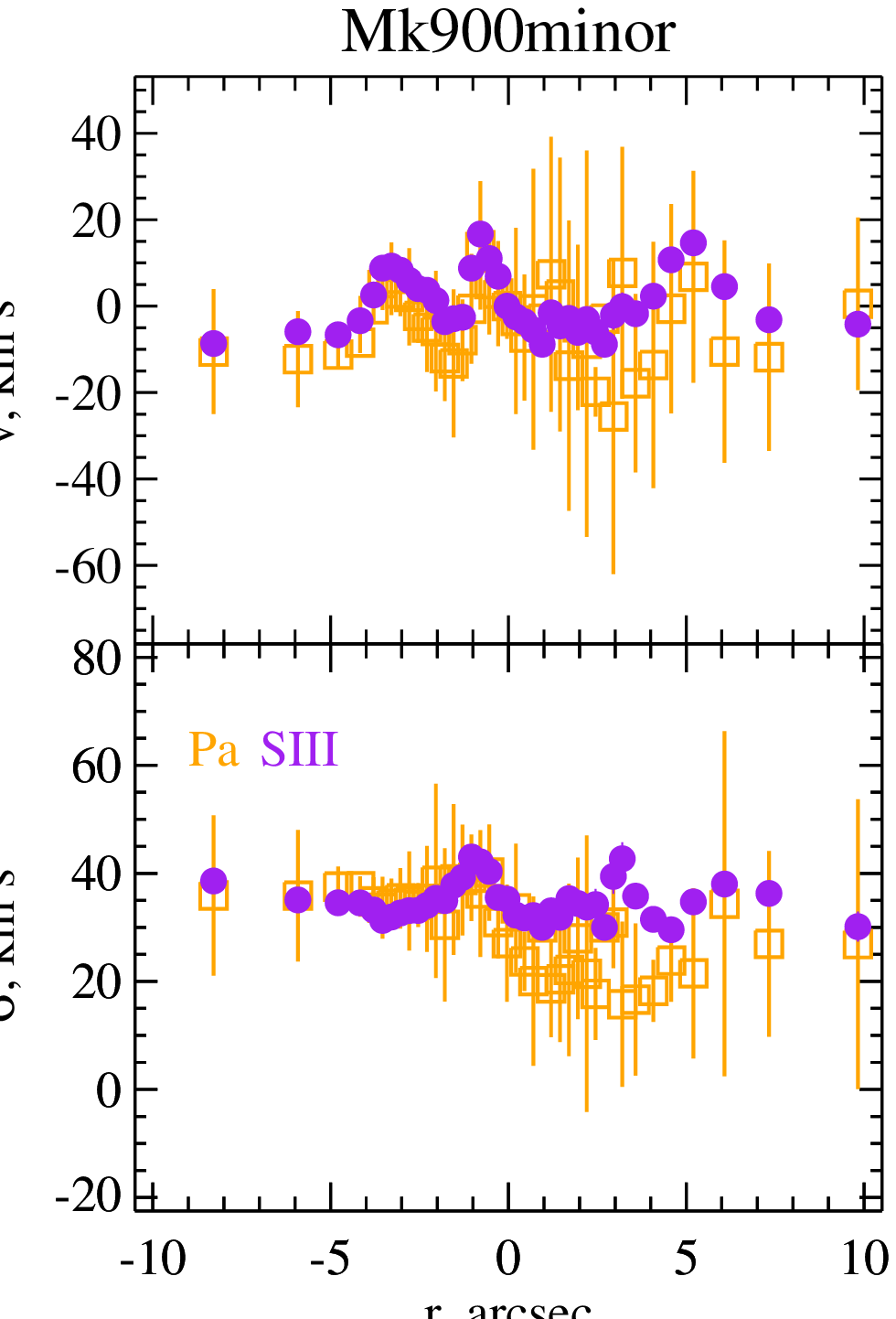}
\includegraphics[width=0.24\textwidth]{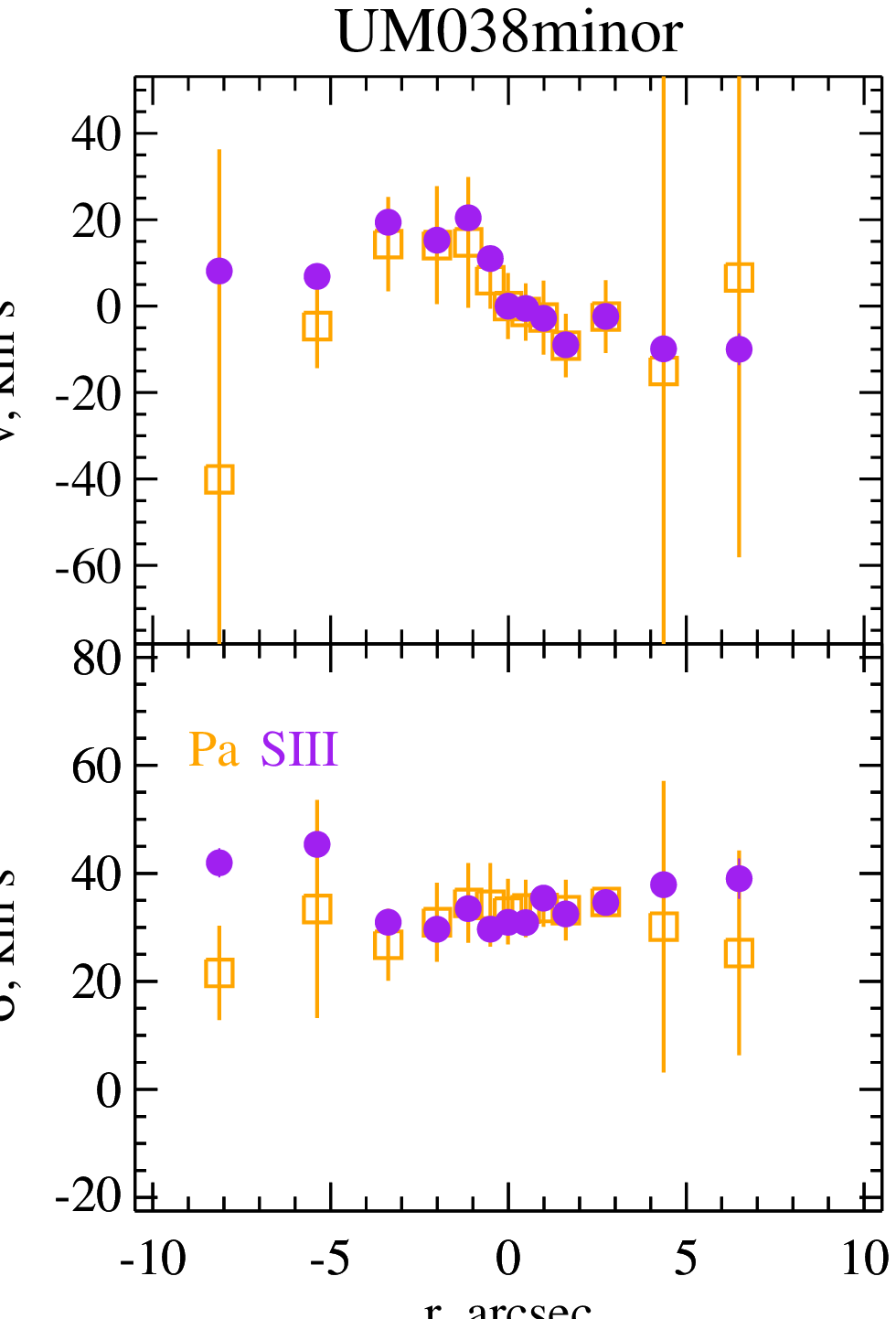}
\includegraphics[width=0.24\textwidth]{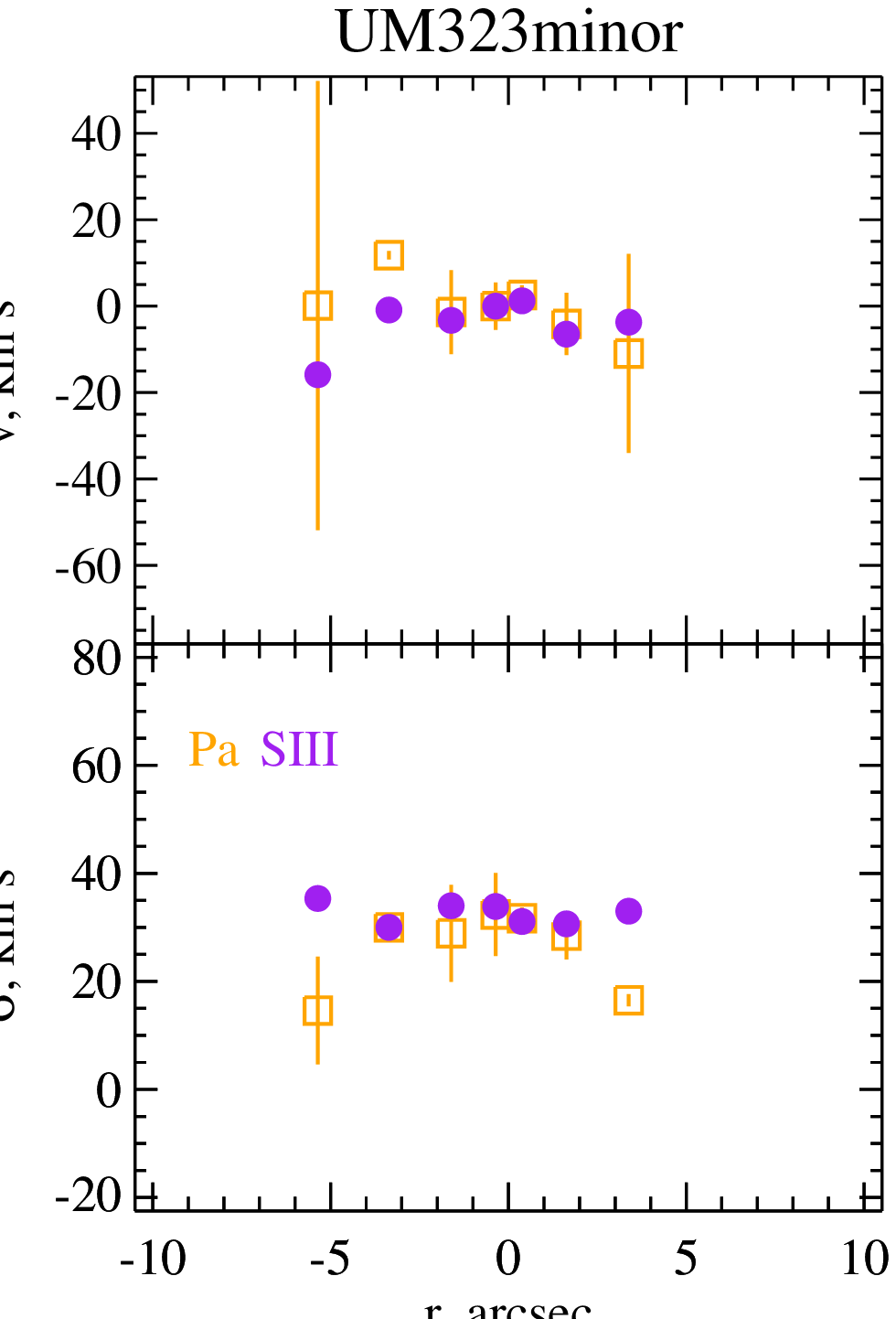}
\caption {Ionized gas velocity fields (upper part
of each panel) and dispersion profiles (lower panel of each panel) for the
major (upper row) and the minor (lower row) axes of our galaxies. With orange
squares we plot the mean values from the Paschen lines. The errors, in
this case, are the standard deviations from the mean. With
purple circles we plot the values derived for the [\sthree 9069]
line. The names of the galaxies and the axis orientations are
indicated on each panel. }
\label{fig:gas}
\end{figure*}

In this section we present the derived kinematical properties of the
minor and major axes of our galaxies (Fig.\,\ref{fig:gas}). The data
were obtained by fitting independent Gaussians to the Paschen (Pa) and
\sthree\ emission lines. The results from the different line fits agree.

\begin{figure*}
\includegraphics[width=0.49\textwidth]{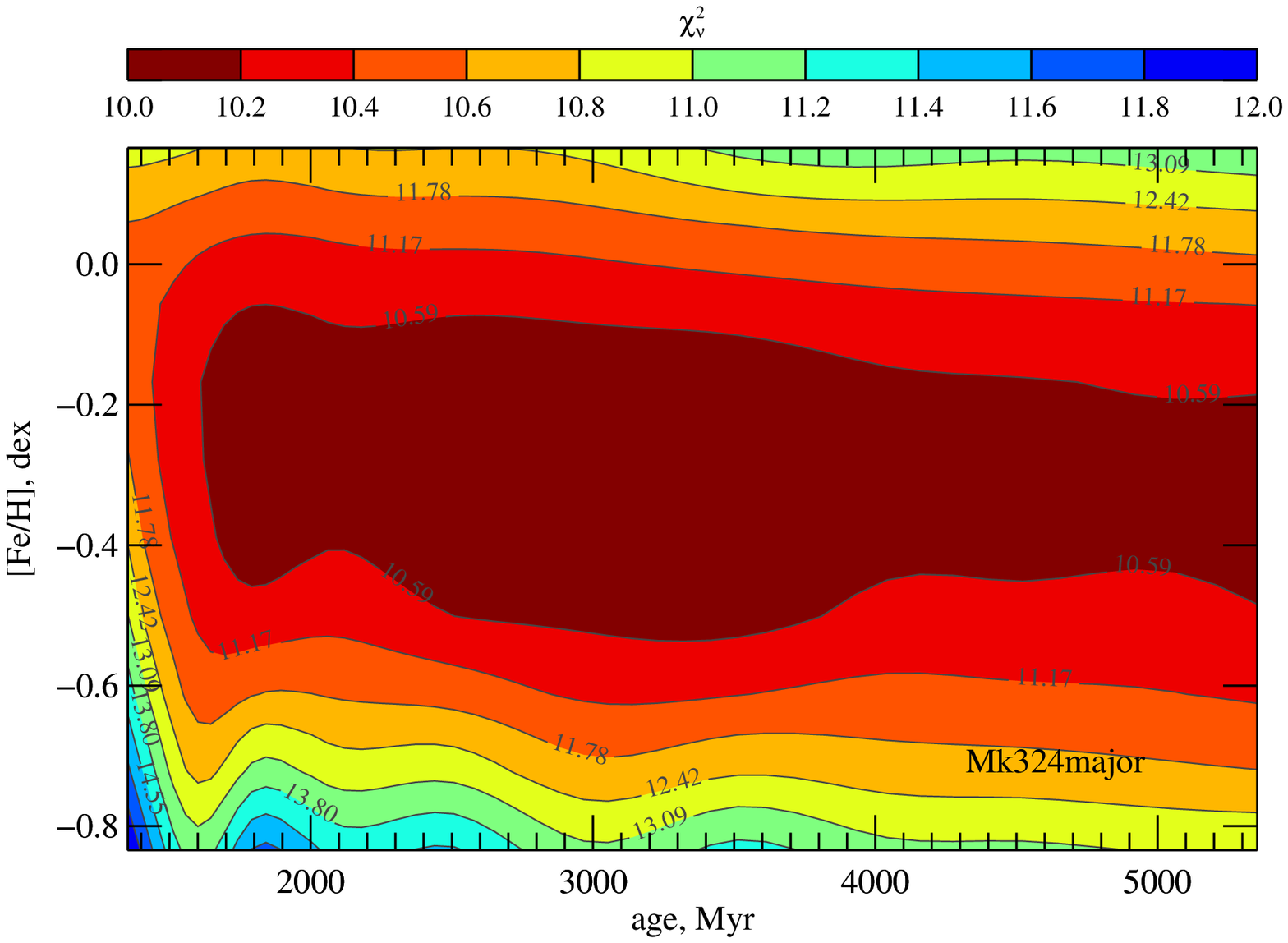}
\includegraphics[width=0.49\textwidth]{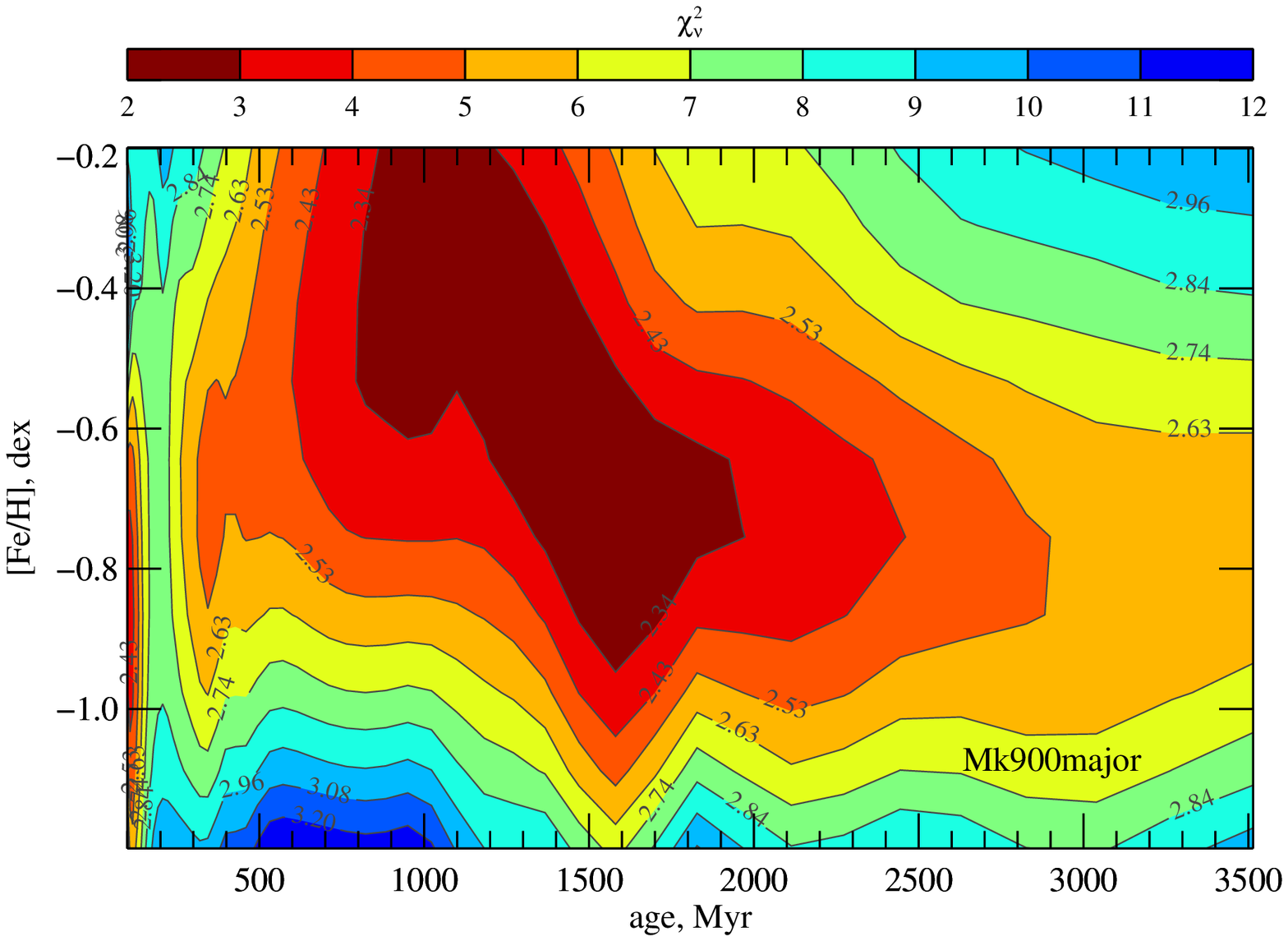}
\includegraphics[width=0.49\textwidth]{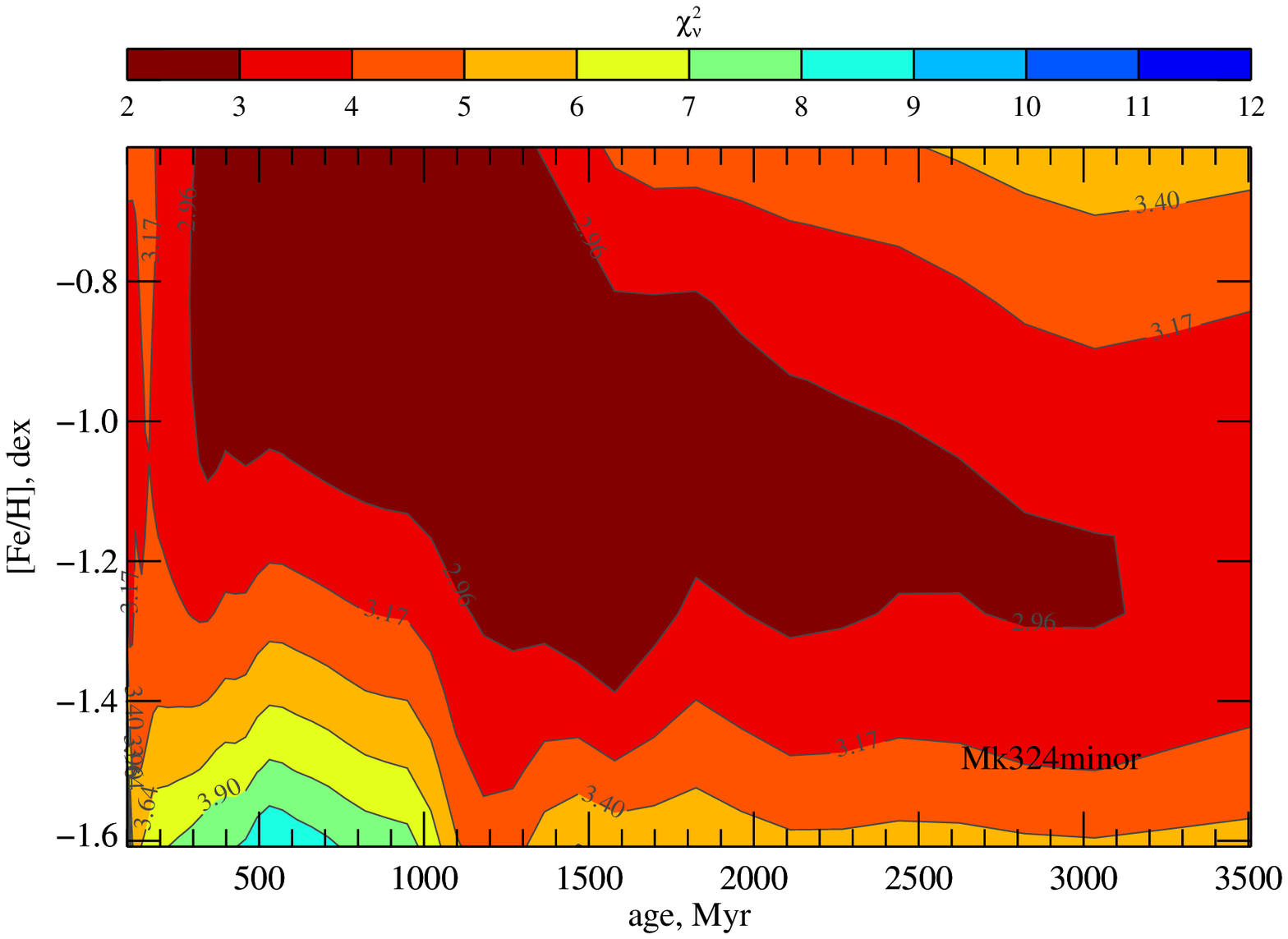}
\includegraphics[width=0.49\textwidth]{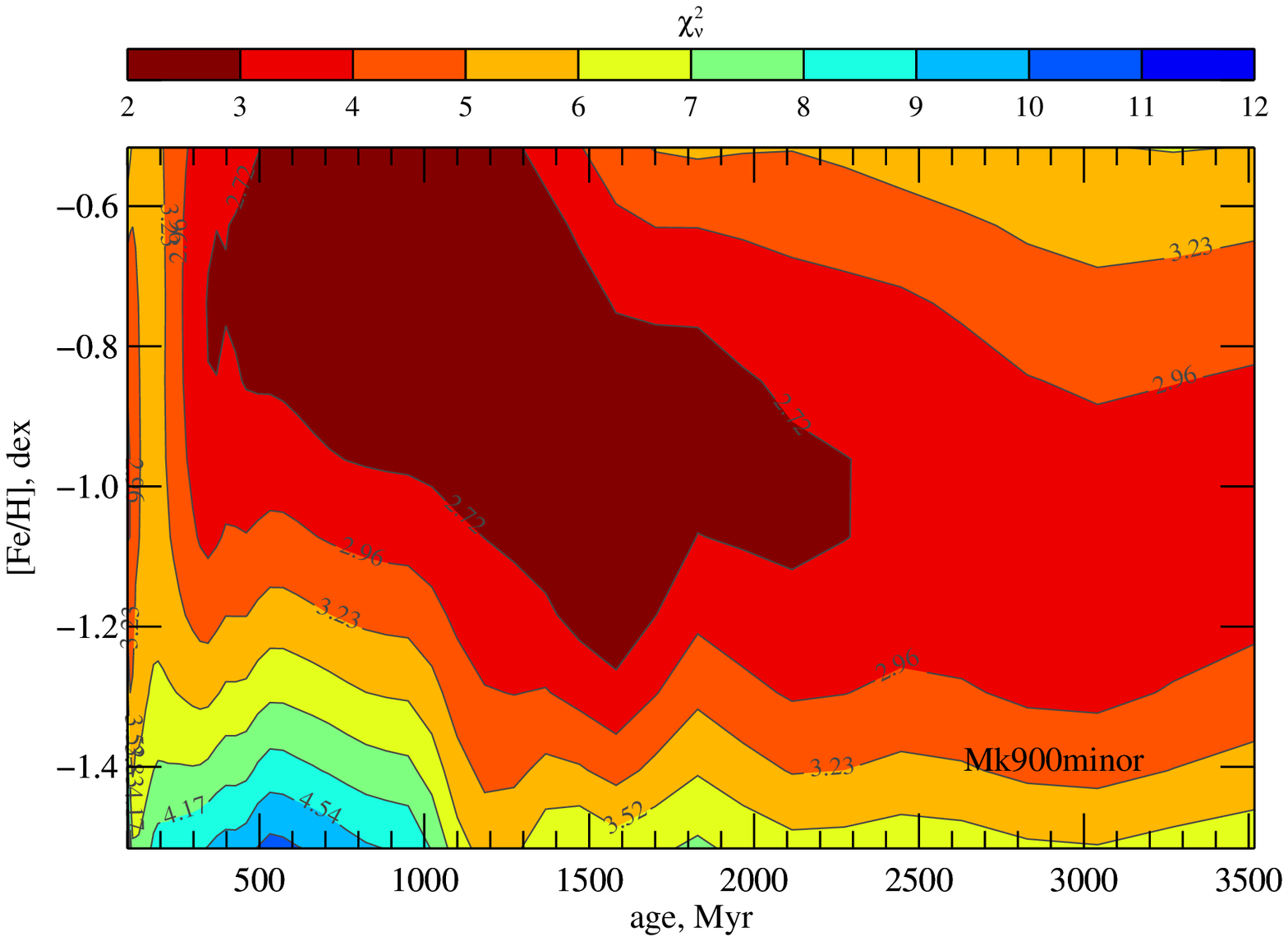}
\includegraphics[width=0.49\textwidth]{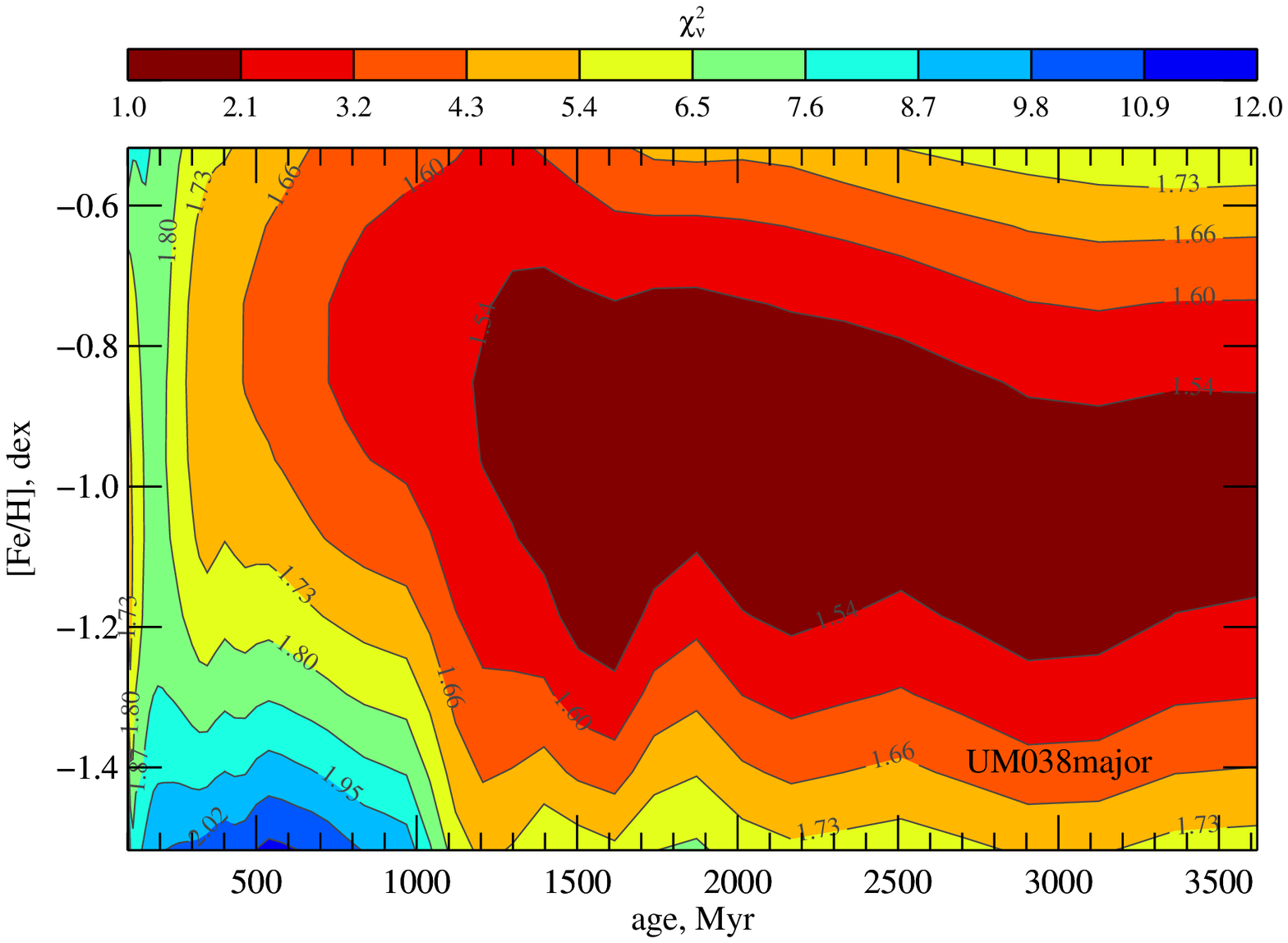}
\includegraphics[width=0.49\textwidth]{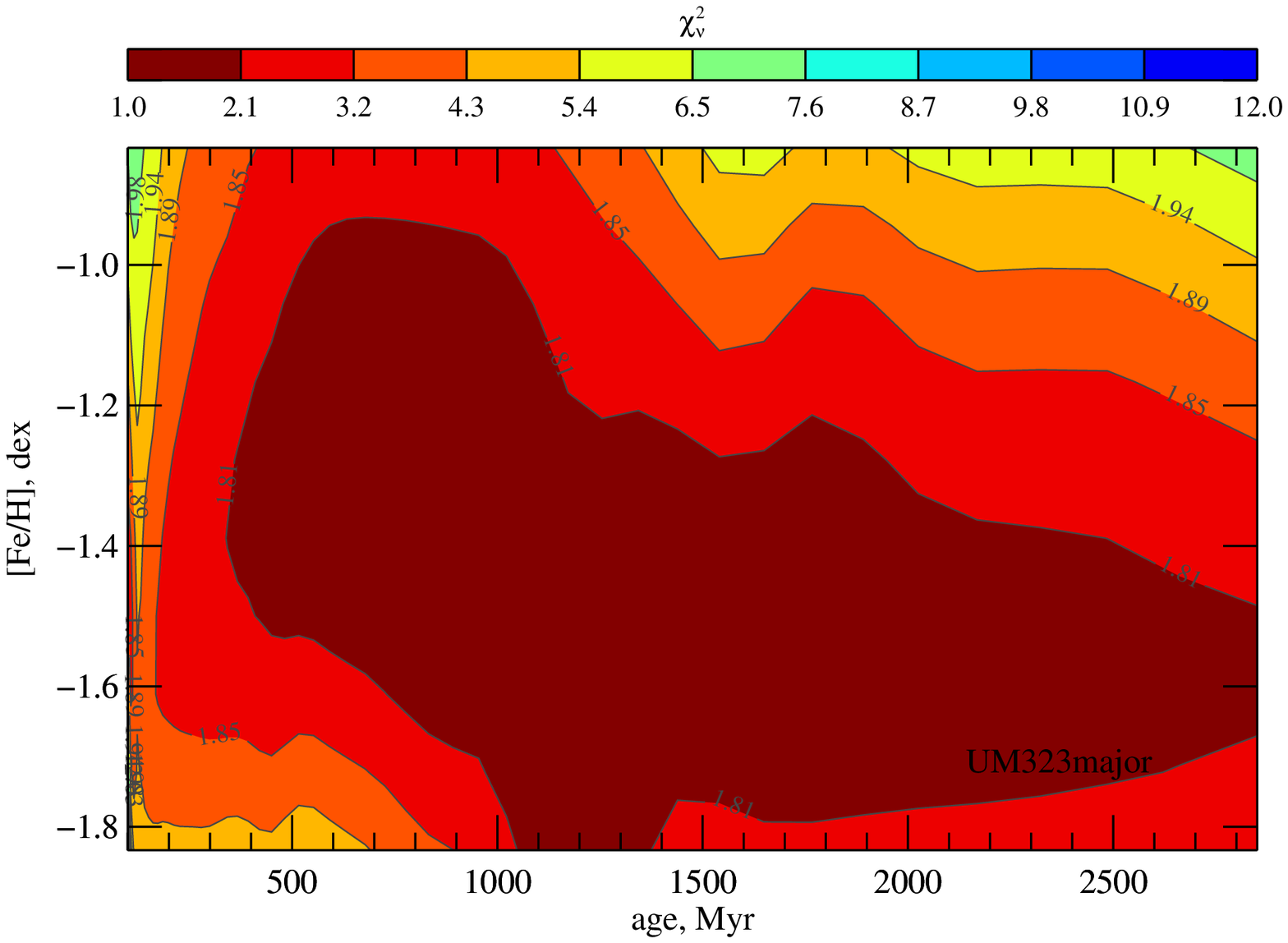}
\includegraphics[width=0.49\textwidth]{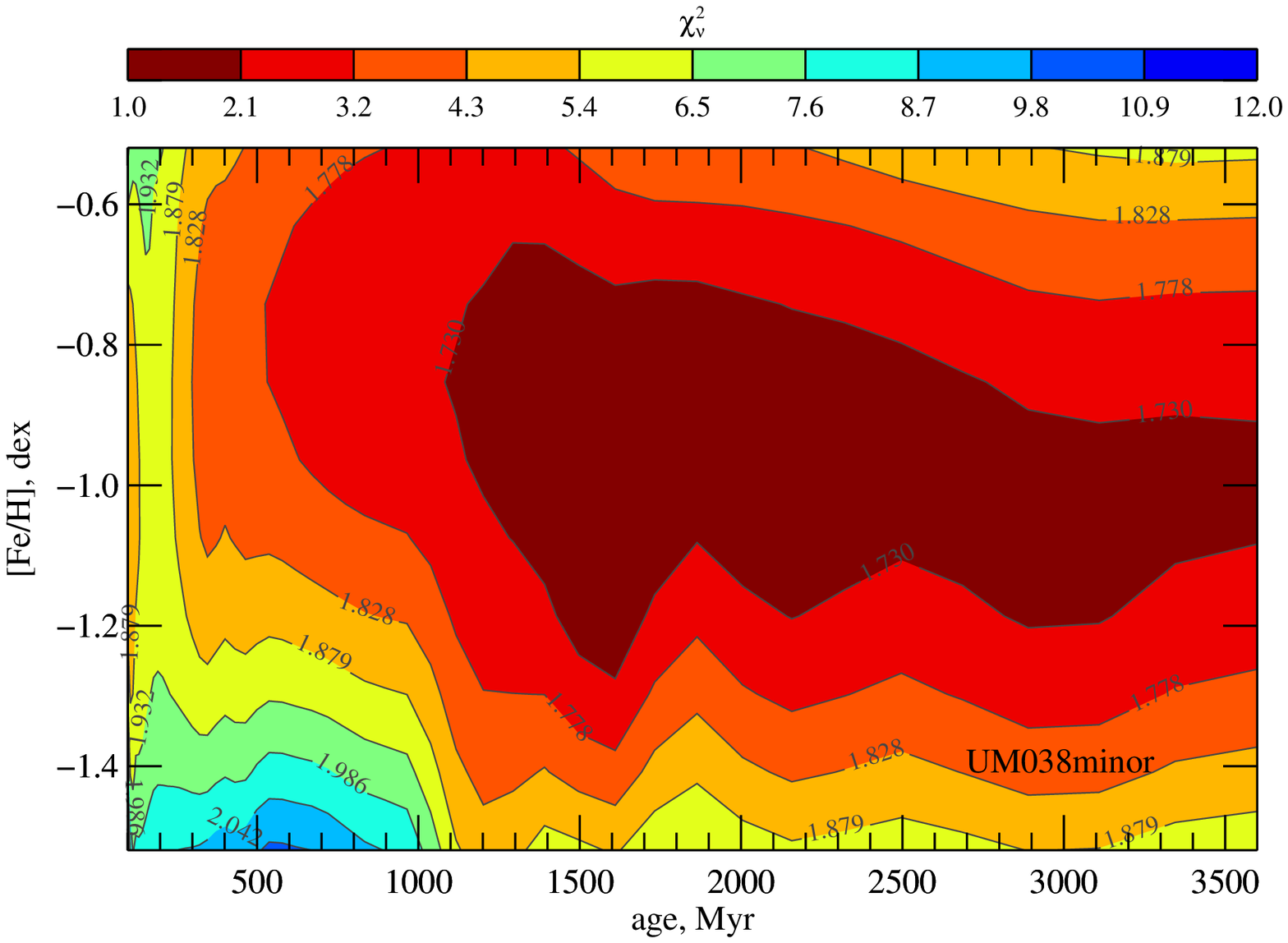}
\includegraphics[width=0.49\textwidth]{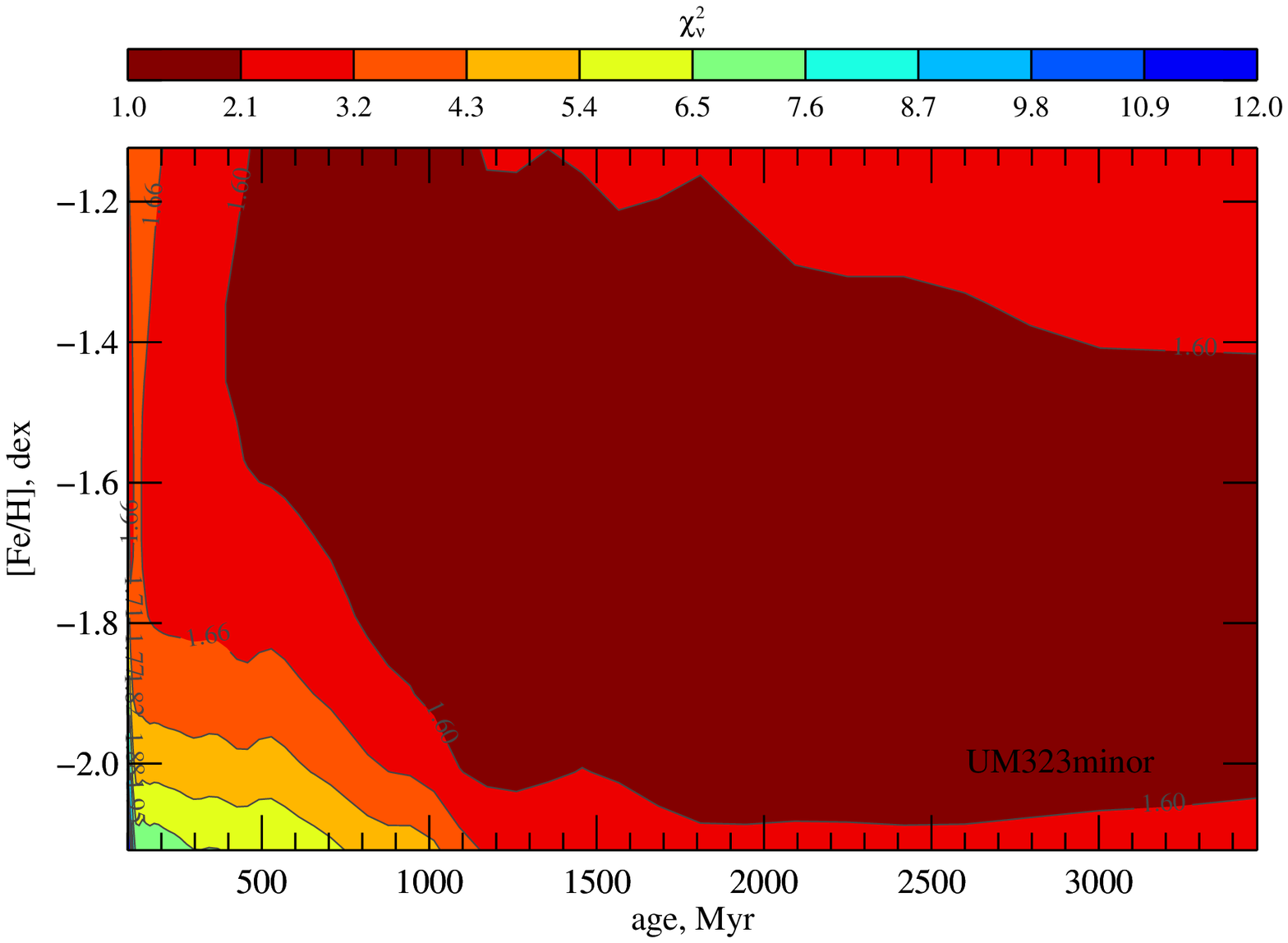}
\caption {$\chi^2$-maps at the luminosity peak for both minor and major
  axes. The maps were constructed by fixing a range of ages and metallicities and
computing the $\chi^2$ value (colour coded). The galaxy names and slit
orientations are indicated on each panel.}
\label{fig:chimaps}
\end{figure*}

Here, we also attach the $\chi^2$-maps at the most prominent
star-formation regions (Fig.\ref{fig:chimaps}) at the minor and major
axes for each of the galaxies.  These maps illustrate the age and
metallicity sensitivity of our spectra. It is clear that we have
little information for stellar ages older than 2\,Gyr. The
metallicities are usually better constrained with an average uncertainty
of 0.2\,dex.

\clearpage

\label{lastpage}

\end{document}

%% file: phot_prop.tex
Mk324 & 15.59 & -16.35 &  3.24 &  0.38 & 14.00 & -17.94 &  5.55 &  0.66\\
Mk900 & 14.59 & -16.69 &  6.90 &  0.60 & 13.11 & -18.17 &  8.14 &  0.71\\
UM038 & 15.91 & -15.63 &  4.66 &  0.46 & 14.80 & -16.74 &  4.67 &  0.46\\
UM323 & 16.12 & -16.01 &  4.48 &  0.58 & 15.09 & -17.04 &  4.78 &  0.62\\

%% file: table_spec.tex
Mk324 & 41.3 & 24.3 & 27.67 $\pm$  5.14 &  1.26 $\pm$  0.38 & -0.83 $\pm$  0.29&0.56&2.79\\
Mk900 & 71.4 & 12.0 & 44.05 $\pm$  5.33 &  1.39 $\pm$  0.37 & -0.88 $\pm$  0.18&0.53&0.35\\
UM038 & 34.8 & 20.4 & 28.69 $\pm$  6.56 &  1.26 $\pm$  0.35 & -1.16 $\pm$  0.17&0.45&1.93\\
UM323 & 49.6 & 21.8 & 36.20 $\pm$ 10.98 &  1.19 $\pm$  0.59 & -1.58 $\pm$  0.36&0.40&1.07\\

%% file: dyn_mass.tex
Mk324& 0.55&  3.16&  2.43&0.76\\
Mk900& 0.54&  6.47&  6.08&1.54\\
UM038& 0.38&  1.68&  1.82&1.72\\
UM323& 0.44&  2.61&  3.32&2.36\\

%% file: bcd.bbl
\begin{thebibliography}{53}
\expandafter\ifx\csname natexlab\endcsname\relax\def\natexlab#1{#1}\fi

\bibitem[{{Bekki}(2008)}]{bekki2008}
{Bekki} K., 2008, \mnras, 388, L10

\bibitem[{{Bigiel} {et~al.}(2010){Bigiel}, {Leroy}, {Walter}, {Blitz},
  {Brinks}, {de Blok}, \& {Madore}}]{bigiel2010}
{Bigiel} F., {Leroy} A., {Walter} F., {Blitz} L., {Brinks} E., {de Blok}
  W.~J.~G., {Madore} B., 2010, \aj, 140, 1194

\bibitem[{{Binney} \& {Tremaine}(2008)}]{binney2008}
{Binney} J., {Tremaine} S., 2008, {Galactic Dynamics: Second Edition}.
  Princeton University Press

\bibitem[{{Chilingarian} {et~al.}(2007){Chilingarian}, {Prugniel},
  {Sil'Chenko}, \& {Afanasiev}}]{chilingarian2007}
{Chilingarian} I.~V., {Prugniel} P., {Sil'Chenko} O.~K., {Afanasiev} V.~L.,
  2007, \mnras, 376, 1033

\bibitem[{{Cloet-Osselaer} {et~al.}(2014){Cloet-Osselaer}, {de Rijcke},
  {Vandenbroucke}, {Schroyen}, \& {Koleva}}]{cloet-osselaer2014}
{Cloet-Osselaer} A., {de Rijcke} S., {Vandenbroucke} B., {Schroyen} J.,
  {Koleva} M., 2014, \mnras, submitted

\bibitem[{{De Rijcke} {et~al.}(2003){De Rijcke}, {Dejonghe}, {Zeilinger}, \&
  {Hau}}]{derijcke2003}
{De Rijcke} S., {Dejonghe} H., {Zeilinger} W.~W., {Hau} G.~K.~T., 2003, \aap,
  400, 119

\bibitem[{{Delisle} \& {Hardy}(1992)}]{delisle1992}
{Delisle} S., {Hardy} E., 1992, \aj, 103, 711

\bibitem[{{Elmegreen} {et~al.}(2009){Elmegreen}, {Elmegreen}, {Fernandez}, \&
  {Lemonias}}]{elmegreen2009}
{Elmegreen} B.~G., {Elmegreen} D.~M., {Fernandez} M.~X., {Lemonias} J.~J.,
  2009, \apj, 692, 12

\bibitem[{{Elmegreen} {et~al.}(2012{\natexlab{a}}){Elmegreen}, {Zhang}, \&
  {Hunter}}]{elemgreen2012}
{Elmegreen} B.~G., {Zhang} H.-X., {Hunter} D.~A., 2012{\natexlab{a}}, \apj,
  747, 105

\bibitem[{{Elmegreen} {et~al.}(2012{\natexlab{b}}){Elmegreen}, {Zhang}, \&
  {Hunter}}]{elmegreen2012}
---, 2012{\natexlab{b}}, \apj, 747, 105

\bibitem[{{Geha} {et~al.}(2005){Geha}, {Guhathakurta}, \& {van der
  Marel}}]{geha2005}
{Geha} M., {Guhathakurta} P., {van der Marel} R.~P., 2005, \aj, 129, 2617

\bibitem[{{Gerola} {et~al.}(1980){Gerola}, {Seiden}, \&
  {Schulman}}]{gerola1980}
{Gerola} H., {Seiden} P.~E., {Schulman} L.~S., 1980, \apj, 242, 517

\bibitem[{{Gil de Paz} {et~al.}(2003){Gil de Paz}, {Madore}, \&
  {Pevunova}}]{gildepaz2003}
{Gil de Paz} A., {Madore} B.~F., {Pevunova} O., 2003, \apjs, 147, 29

\bibitem[{{Goerdt} {et~al.}(2010){Goerdt}, {Moore}, {Read}, \&
  {Stadel}}]{goerdt2010}
{Goerdt} T., {Moore} B., {Read} J.~I., {Stadel} J., 2010, \apj, 725, 1707

\bibitem[{{Gordon} \& {Gottesman}(1981)}]{gordon1981}
{Gordon} D., {Gottesman} S.~T., 1981, \aj, 86, 161

\bibitem[{{Hinz} {et~al.}(2001){Hinz}, {Rix}, \& {Bernstein}}]{hinz2001}
{Hinz} J.~L., {Rix} H.-W., {Bernstein} G.~M., 2001, \aj, 121, 683

\bibitem[{{Hunter} \& {Elmegreen}(2004)}]{hunter2004}
{Hunter} D.~A., {Elmegreen} B.~G., 2004, \aj, 128, 2170

\bibitem[{{Hunter} \& {Elmegreen}(2006)}]{hunter2006}
---, 2006, \apjs, 162, 49

\bibitem[{{Hunter} {et~al.}(2012){Hunter}, {Ficut-Vicas}, {Ashley}, {Brinks},
  {Cigan}, {Elmegreen}, {Heesen}, {Herrmann}, {Johnson}, {Oh}, {Rupen},
  {Schruba}, {Simpson}, {Walter}, {Westpfahl}, {Young}, \&
  {Zhang}}]{hunter2012}
{Hunter} D.~A., {Ficut-Vicas} D., {Ashley} T., {Brinks} E., {Cigan} P.,
  {Elmegreen} B.~G., {Heesen} V., {Herrmann} K.~A., {Johnson} M., {Oh} S.-H.,
  {Rupen} M.~P., {Schruba} A., {Simpson} C.~E., {Walter} F., {Westpfahl} D.~J.,
  {Young} L.~M., {Zhang} H.-X., 2012, \aj, 144, 134

\bibitem[{{Kelson}(2003)}]{kelson2003}
{Kelson} D.~D., 2003, \pasp, 115, 688

\bibitem[{{Koleva} {et~al.}(2013){Koleva}, {Bouchard}, {Prugniel}, {De Rijcke},
  \& {Vauglin}}]{koleva2013}
{Koleva} M., {Bouchard} A., {Prugniel} P., {De Rijcke} S., {Vauglin} I., 2013,
  \mnras, 428, 2949

\bibitem[{{Koleva} {et~al.}(2009{\natexlab{a}}){Koleva}, {de Rijcke},
  {Prugniel}, {Zeilinger}, \& {Michielsen}}]{koleva2009}
{Koleva} M., {de Rijcke} S., {Prugniel} P., {Zeilinger} W.~W., {Michielsen} D.,
  2009{\natexlab{a}}, \mnras, 396, 2133

\bibitem[{{Koleva} {et~al.}(2009{\natexlab{b}}){Koleva}, {Prugniel},
  {Bouchard}, \& {Wu}}]{ulyss}
{Koleva} M., {Prugniel} P., {Bouchard} A., {Wu} Y., 2009{\natexlab{b}}, \aap,
  501, 1269

\bibitem[{{Koleva} {et~al.}(2011){Koleva}, {Prugniel}, {de Rijcke}, \&
  {Zeilinger}}]{koleva2011}
{Koleva} M., {Prugniel} P., {de Rijcke} S., {Zeilinger} W.~W., 2011, \mnras,
  417, 1643

\bibitem[{{Koleva} \& {Vazdekis}(2012)}]{koleva2012}
{Koleva} M., {Vazdekis} A., 2012, \aap, 538, A143

\bibitem[{{Kunth} {et~al.}(1988){Kunth}, {Maurogordato}, \&
  {Vigroux}}]{kunth1988}
{Kunth} D., {Maurogordato} S., {Vigroux} L., 1988, \aap, 204, 10

\bibitem[{{Le Borgne} {et~al.}(2004){Le Borgne}, {Rocca-Volmerange},
  {Prugniel}, {Lan{\c c}on}, {Fioc}, \& {Soubiran}}]{pegase.hr}
{Le Borgne} D., {Rocca-Volmerange} B., {Prugniel} P., {Lan{\c c}on} A., {Fioc}
  M., {Soubiran} C., 2004, \aap, 425, 881

\bibitem[{{Lelli} {et~al.}(2013){Lelli}, {Fraternali}, \&
  {Verheijen}}]{lelli2013}
{Lelli} F., {Fraternali} F., {Verheijen} M., 2013, ArXiv e-prints

\bibitem[{{Lisker} {et~al.}(2006){Lisker}, {Glatt}, {Westera}, \&
  {Grebel}}]{lisker2006}
{Lisker} T., {Glatt} K., {Westera} P., {Grebel} E.~K., 2006, \aj, 132, 2432

\bibitem[{{Loose} \& {Thuan}(1986)}]{loose1986}
{Loose} H.-H., {Thuan} T.~X., 1986, in Star-forming Dwarf Galaxies and Related
  Objects, {Kunth} D., {Thuan} T.~X., {Tran Thanh Van} J., {Lequeux} J.,
  {Audouze} J., eds., pp. 73--88

\bibitem[{{Mart{\'{\i}}nez-Delgado} {et~al.}(2007){Mart{\'{\i}}nez-Delgado},
  {Tenorio-Tagle}, {Mu{\~n}oz-Tu{\~n}{\'o}n}, {Moiseev}, \&
  {Cair{\'o}s}}]{martinez2007}
{Mart{\'{\i}}nez-Delgado} I., {Tenorio-Tagle} G., {Mu{\~n}oz-Tu{\~n}{\'o}n} C.,
  {Moiseev} A.~V., {Cair{\'o}s} L.~M., 2007, \aj, 133, 2892

\bibitem[{{Meyer} {et~al.}(2013){Meyer}, {Lisker}, {Janz}, \&
  {Papaderos}}]{meyer2013}
{Meyer} H.~T., {Lisker} T., {Janz} J., {Papaderos} P., 2013, ArXiv e-prints

\bibitem[{{{\"O}stlin} {et~al.}(2004){{\"O}stlin}, {Cumming}, {Amram},
  {Bergvall}, {Kunth}, {M{\'a}rquez}, {Masegosa}, \& {Zackrisson}}]{ostlin2004}
{{\"O}stlin} G., {Cumming} R.~J., {Amram} P., {Bergvall} N., {Kunth} D.,
  {M{\'a}rquez} I., {Masegosa} J., {Zackrisson} E., 2004, \aap, 419, L43

\bibitem[{{Paturel} {et~al.}(2003){Paturel}, {Petit}, {Prugniel}, {Theureau},
  {Rousseau}, {Brouty}, {Dubois}, \& {Cambr{\'e}sy}}]{hyperleda}
{Paturel} G., {Petit} C., {Prugniel} P., {Theureau} G., {Rousseau} J., {Brouty}
  M., {Dubois} P., {Cambr{\'e}sy} L., 2003, \aap, 412, 45

\bibitem[{{Prugniel} \& {Simien}(1997)}]{prugniel1997}
{Prugniel} P., {Simien} F., 1997, \aap, 321, 111

\bibitem[{{Prugniel} \& {Simien}(2003)}]{prugniel2003}
---, 2003, \apss, 284, 603

\bibitem[{{Prugniel} {et~al.}(2011){Prugniel}, {Vauglin}, \&
  {Koleva}}]{prugniel2011}
{Prugniel} P., {Vauglin} I., {Koleva} M., 2011, \aap, 531, A165

\bibitem[{{Ry{\'s}} {et~al.}(2013){Ry{\'s}}, {Falc{\'o}n-Barroso}, \& {van de
  Ven}}]{rys2013}
{Ry{\'s}} A., {Falc{\'o}n-Barroso} J., {van de Ven} G., 2013, \mnras, 428, 2980

\bibitem[{{S{\'a}nchez Almeida} {et~al.}(2008){S{\'a}nchez Almeida},
  {Mu{\~n}oz-Tu{\~n}{\'o}n}, {Amor{\'{\i}}n}, {Aguerri}, {S{\'a}nchez-Janssen},
  \& {Tenorio-Tagle}}]{sanchez-almeida2008}
{S{\'a}nchez Almeida} J., {Mu{\~n}oz-Tu{\~n}{\'o}n} C., {Amor{\'{\i}}n} R.,
  {Aguerri} J.~A., {S{\'a}nchez-Janssen} R., {Tenorio-Tagle} G., 2008, \apj,
  685, 194

\bibitem[{{Schroyen} {et~al.}(2013){Schroyen}, {De Rijcke}, {Koleva},
  {Cloet-Osselaer}, \& {Vandenbroucke}}]{schroyen2013}
{Schroyen} J., {De Rijcke} S., {Koleva} M., {Cloet-Osselaer} A.,
  {Vandenbroucke} B., 2013, \mnras, 434, 888

\bibitem[{{Toloba} {et~al.}(2011){Toloba}, {Boselli}, {Cenarro}, {Peletier},
  {Gorgas}, {Gil de Paz}, \& {Mu{\~n}oz-Mateos}}]{toloba2011}
{Toloba} E., {Boselli} A., {Cenarro} A.~J., {Peletier} R.~F., {Gorgas} J., {Gil
  de Paz} A., {Mu{\~n}oz-Mateos} J.~C., 2011, \aap, 526, A114

\bibitem[{{Tremonti} {et~al.}(2004){Tremonti}, {Heckman}, {Kauffmann},
  {Brinchmann}, {Charlot}, {White}, {Seibert}, {Peng}, {Schlegel}, {Uomoto},
  {Fukugita}, \& {Brinkmann}}]{tremonti2004}
{Tremonti} C.~A., {Heckman} T.~M., {Kauffmann} G., {Brinchmann} J., {Charlot}
  S., {White} S.~D.~M., {Seibert} M., {Peng} E.~W., {Schlegel} D.~J., {Uomoto}
  A., {Fukugita} M., {Brinkmann} J., 2004, \apj, 613, 898

\bibitem[{{Valdes} {et~al.}(2004){Valdes}, {Gupta}, {Rose}, {Singh}, \&
  {Bell}}]{valdes2004}
{Valdes} F., {Gupta} R., {Rose} J.~A., {Singh} H.~P., {Bell} D.~J., 2004,
  \apjs, 152, 251

\bibitem[{{van Zee} {et~al.}(2001){van Zee}, {Salzer}, \&
  {Skillman}}]{vanzee2001}
{van Zee} L., {Salzer} J.~J., {Skillman} E.~D., 2001, \aj, 122, 121

\bibitem[{{van Zee} {et~al.}(1998){van Zee}, {Skillman}, \&
  {Salzer}}]{vanzee1998}
{van Zee} L., {Skillman} E.~D., {Salzer} J.~J., 1998, \aj, 116, 1186

\bibitem[{{Vazdekis} {et~al.}(2003){Vazdekis}, {Cenarro}, {Gorgas}, {Cardiel},
  \& {Peletier}}]{vazdekis2003}
{Vazdekis} A., {Cenarro} A.~J., {Gorgas} J., {Cardiel} N., {Peletier} R.~F.,
  2003, \mnras, 340, 1317

\bibitem[{{Verbeke} {et~al.}(in prep.){Verbeke}, {de Rijcke}, {Koleva},
  {Cloet-Osselaer}, \& {Vandenbroucke}}]{verbeke2014}
{Verbeke} R., {de Rijcke} S., {Koleva} M., {Cloet-Osselaer} A., {Vandenbroucke}
  B., in prep.

\bibitem[{{Wakker} \& {van Woerden}(1997)}]{wakker1997}
{Wakker} B.~P., {van Woerden} H., 1997, \araa, 35, 217

\bibitem[{{Walker} {et~al.}(2009){Walker}, {Mateo}, {Olszewski},
  {Pe{\~n}arrubia}, {Wyn Evans}, \& {Gilmore}}]{walker2009}
{Walker} M.~G., {Mateo} M., {Olszewski} E.~W., {Pe{\~n}arrubia} J., {Wyn Evans}
  N., {Gilmore} G., 2009, \apj, 704, 1274

\bibitem[{{Weisz} {et~al.}(2011){Weisz}, {Dalcanton}, {Williams}, {Gilbert},
  {Skillman}, {Seth}, {Dolphin}, {McQuinn}, {Gogarten}, {Holtzman}, {Rosema},
  {Cole}, {Karachentsev}, \& {Zaritsky}}]{weisz2011}
{Weisz} D.~R., {Dalcanton} J.~J., {Williams} B.~F., {Gilbert} K.~M., {Skillman}
  E.~D., {Seth} A.~C., {Dolphin} A.~E., {McQuinn} K.~B.~W., {Gogarten} S.~M.,
  {Holtzman} J., {Rosema} K., {Cole} A., {Karachentsev} I.~D., {Zaritsky} D.,
  2011, \apj, 739, 5

\bibitem[{{Worthey}(1999)}]{worthey1999}
{Worthey} G., 1999, in Astronomical Society of the Pacific Conference Series,
  Vol. 192, Spectrophotometric Dating of Stars and Galaxies, {Hubeny} I.,
  {Heap} S., {Cornett} R., eds., p. 283

\bibitem[{{Wu} {et~al.}(2011){Wu}, {Singh}, {Prugniel}, {Gupta}, \&
  {Koleva}}]{wu2011}
{Wu} Y., {Singh} H.~P., {Prugniel} P., {Gupta} R., {Koleva} M., 2011, \aap,
  525, A71

\bibitem[{{Zhao} {et~al.}(2013){Zhao}, {Gao}, \& {Gu}}]{zhao2013}
{Zhao} Y., {Gao} Y., {Gu} Q., 2013, \apj, 764, 44

\end{thebibliography}
